\documentclass[aps,prb,reprint,groupedaddress,amsmath,amssymb,superscriptaddress]{revtex4-2}
\usepackage{graphicx}
\usepackage{float}
\usepackage{multirow}
\usepackage{natbib,twoopt}
\usepackage[breaklinks=true]{hyperref}
\usepackage{xcolor}
\usepackage{amssymb}
\usepackage{stackengine,graphicx}
\usepackage{amsmath}
\usepackage{hyperref}
\usepackage{color}
\usepackage{stackengine}
\usepackage{subfigure}
\usepackage{verbatim}
\usepackage{mathtools}
\usepackage{braket}
\usepackage{bm}

\newcommand{\rot}{\mathop{\rm rot}\nolimits}
\newcommand{\divv}{\mathop{\rm div}\nolimits}
\newcommand{\df}[2]{\frac{\partial #1}{\partial #2}}

\newcommand{\eps}{\varepsilon}

\newcommand{\cd}{\tilde{\chi}}


\begin{document}

\title{Dual origin of effective axion response}

\author{Timur Z. Seidov}
\affiliation{School of Physics and Engineering, ITMO University, Saint  Petersburg, Russia}

\author{Eduardo Barredo-Alamilla}
\affiliation{School of Physics and Engineering, ITMO University, Saint  Petersburg, Russia}

\author{Daniel A. Bobylev}
\affiliation{School of Physics and Engineering, ITMO University, Saint  Petersburg, Russia}

\author{Leon Shaposhnikov}
\affiliation{School of Physics and Engineering, ITMO University, Saint  Petersburg, Russia}

\author{Maxim Mazanov}
\affiliation{School of Physics and Engineering, ITMO University, Saint Petersburg, Russia}

\author{Maxim A. Gorlach}
\email{m.gorlach@metalab.ifmo.ru}
\affiliation{School of Physics and Engineering, ITMO University, Saint Petersburg, Russia}

\begin{abstract}
Effective axion fields in condensed matter and photonics are manifested as $\mathcal{P}$- and $\mathcal{T}$-odd contributions to the electromagnetic response. Here, we show that the phenomena previously attributed to the effective axion fields have two distinct physical origins. One of them corresponds to the standard axion electrodynamics, while another provides its dual-symmetric version having the same symmetry and featuring similar but distinguishable optical properties. We present an example system described by the dual-symmetric modification of axion electrodynamics, derive the key predictions and pinpoint experimentally observable distinctions between the two versions of axion-type response.
\end{abstract}

\date{\today}

\maketitle



Effective description of condensed matter and photonic structures allows to reduce their complex behavior to relatively simple theoretical models ignoring redundant degrees of freedom and introducing several key parameters. This approach features profound analogies with quantum field theory. For instance, physics of relativistic Dirac fermions is reproduced in graphene, condensed matter and photonic systems described by the Dirac effective Hamiltonian~\cite{Bernevig2006,Raghu2008,Shen}. Quantum Hall effect can be concisely described using the effective Chern-Simons Lagrangian~\cite{Tong2016,Greiter2022}, and the physics of topological insulators including quadrupole insulators is  captured by the field-theoretical language~\cite{Qi2008,You2021}.

Given their versatility, condensed matter and photonic structures can be a source of various effective theories, sometimes even without a known counterpart within the Standard Model. In this spirit, hypothetic axions~\cite{Wilczek78,Weinberg78} that evaded their experimental observation so far, can be realized as collective excitations in a class of condensed matter~\cite{Li2010,Wu2016,Nenno2020,Sekine2021} and photonic systems~\cite{Shaposhnikov2023,Asadchy2024,ShuangZhang24}.

The presence of effective axion fields modifies the equations of electromagnetism which can be presented in the form~\cite{Shaposhnikov2023}
\begin{gather}
\rot\left( -\chi\,{\bf E}+{\bf H}\right)=\frac{1}{c}\,\df{ }{t}\,\left({\bf D}+\chi\,{\bf B}\right)+\frac{4\pi}{c}\,{\bf j}\:,\notag\\
\divv\left({\bf D}+\chi\,{\bf B}\right)=4\pi\rho\:,\label{eq:Chi1}\\
\rot{\bf E}=-\frac{1}{c}\,\df{{\bf B}}{t}\:,\mspace{8mu}\divv{\bf B}=0\:.\label{eq:Chi2}
\end{gather}
Here $\rho$ and ${\bf j}$ are the external charges and currents, ${\bf D}=\eps\,{\bf E}$, ${\bf B}=\mu\,{\bf H}$, while $\eps$ and $\mu$ are permittivity and permeability of the medium, respectively.

The difference from the conventional electrodynamics appears in the first pair of Maxwell's equations Eqs.~\eqref{eq:Chi1} and is rooted in the nonzero $\chi$ which is called {\it effective axion field} in condensed matter literature or {\it Tellegen coefficient} in photonics. This term breaks the inversion $\mathcal{P}$ and time-reversal $\mathcal{T}$ symmetry of the system while keeping $\mathcal{PT}$ symmetry preserved.

Interestingly, if $\chi$ is time-independent ($\partial\chi/\partial t=0$) and spatially homogeneous, as is the case for many condensed matter and photonic systems, it does not affect bulk properties of the medium and arises only at the boundaries. This could be immediately seen from Eqs.~\eqref{eq:Chi1} by recasting them in the form
\begin{gather}
    \rot{\bf H}=\frac{1}{c}\,\df{{\bf D}}{t}+\nabla\chi\times{\bf E}+\frac{4\pi}{c}\,{\bf j}\:,\notag\\
    \divv{\bf D}=-\nabla\chi\cdot{\bf B}+4\pi\rho\:.\label{eq:chi3}
\end{gather}

Hence, $\chi$ is a bulk property of the medium, which manifests itself only at the boundary. Partly due to this subtlety, the very existence of Tellegen media was actively disputed in photonics literature~\cite{Post1965,Lakhtakia1994,Sihvola1995,Tretyakov}. By now, the existence of $\chi$ response is well established in multiple systems including antiferromagnetic structures, magneto-electrics, multiferroics and three-dimensional topological insulators~\cite{Nenno2020,Sekine2021}. In the latter case, the axion field plays the role of topological invariant and is quantized in units of the fine-structure constant $\alpha$ as confirmed experimentally~\cite{Wu2016}.

In condensed matter physics, the axion response is derived through the quantum-mechanical calculation utilizing the notion of Berry connection~\cite{Essin2009} which yields quantized and frequency-independent $\chi$. However, recent studies~\cite{Ahn2022,Wang2022} suggest that this picture is simplified and additional effects arise.



In this Article, we approach this field from a different direction and predict the existence of a novel type of electromagnetic response further termed {\it dual axion field}, $\cd$. Similarly to the conventional axion, $\cd$ manifests itself only at the boundaries and shares with $\chi$ the same parity with respect to the spatial inversion $\mathcal{P}$ and time reversal $\mathcal{T}$. At the same time, the effective theory with $\cd$ leads to the equations of electrodynamics {\it with effective magnetic charge} resulting in the distinct and practically uncharted physics providing a new twist in the effective description of the structured media.

\section*{Heuristics of dual axion response}

While the theory of axion response in condensed matter systems is well established~\cite{Qi2008,Essin2009,Nenno2020,Sekine2021}, its origin in classical photonic structures remained much less studied. Electromagnetic community focused on realizing Tellegen particles~\cite{Tretyakov2003,Asadchy2024}, while the emergence of the collective Tellegen response was poorly understood.  

\begin{figure}[b]
    \centering
    \includegraphics[width=0.95\linewidth]{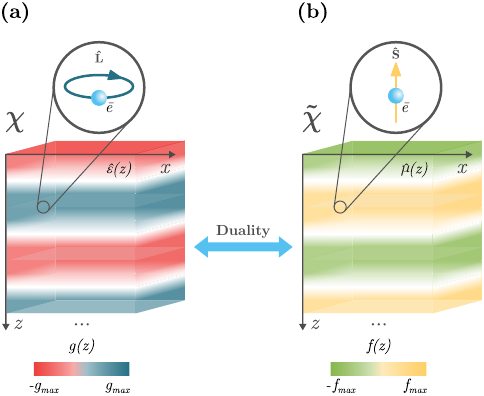}
    \caption{\label{fig:scheme} Schematic of the two structures with broken $\mathcal{P}$ and $\mathcal{T}$ symmetries but invariant under $\mathcal{PT}$. (a) Design of the  metamaterial with effective axion response $\chi$, Ref.~\cite{Shaposhnikov2023}. Gyrotropic permittivity Eq.~\eqref{eq:GEps} is typically associated with the orbital motion, see inset. (b) Metamaterial featuring dual axion response $\tilde{\chi}$. Gyrotropic permeability Eq.~\eqref{eq:Gmu} is typically due to the spin contribution, see inset.}
\end{figure}

A useful insight has recently been provided by Ref.~\cite{Shaposhnikov2023} which suggested a simple photonic structure based on the stack of layers with gyrotropic permittivity 
\begin{equation}\label{eq:GEps}
    \hat{\eps}=\begin{pmatrix}
\eps & ig(z) & 0\\
-ig(z) & \eps & 0\\
0 & 0 & \eps
    \end{pmatrix}\:,
\end{equation}
where out-of-plane magnetization of the layers $g(z)$ is modulated periodically with the zero average [Fig.~\ref{fig:scheme}(a)]. The treatment of this system solely based on Maxwell's equations leads to the 
 axion electrodynamics Eqs.~\eqref{eq:Chi1},\eqref{eq:Chi2} for the averaged fields, where the strength of the effective axion response
\begin{equation}
    \chi=-\frac{2\pi a}{\lambda}\,\int\limits_0^1\,g(sa)\,s\,ds
\end{equation}
is proportional to the period-to-wavelength ratio $a/\lambda$ and depends on the distribution of magnetization $g(z)$ within the unit cell. Interestingly, $\chi$ also depends on the chosen termination of the structure, contrasting with the behavior of the conventional bulk material parameters~\cite{Alu2011}.

Drawing the intuition from this model system, we introduce a similar structure composed of the layers with gyrotropic permeability $\hat{\mu}$
\begin{equation}\label{eq:Gmu}
    \hat{\mu}=\begin{pmatrix}
\mu & if(z) & 0\\
-if(z) & \mu & 0\\
0 & 0 & \mu
    \end{pmatrix}\:,
\end{equation}
and a similar periodic distribution of magnetization $f(z)$ [Fig.~\ref{fig:scheme}(b)]. Clearly, such system shares with axion metamaterial the same symmetries with respect to spatial inversion $\mathcal{P}$ and time reversal $\mathcal{T}$. Moreover, given the microscopic (non-averaged) fields ${\bf e}$, ${\bf d}$, ${\bf b}$ and ${\bf h}$ in an axion metamaterial, one can immediately find the respective fields in our structure performing the dual symmetry transformation:
\begin{align}
  &{\bf e}\rightarrow{\bf h}, \mspace{15mu}{\bf d}\rightarrow{\bf b},  \mspace{15mu} \hat{\eps}\rightarrow\hat{\mu} \notag\\
  &{\bf h}\rightarrow -{\bf e}, \mspace{15mu}{\bf b}\rightarrow -{\bf d},  \mspace{15mu}\hat{\mu}\rightarrow\hat{\eps}\:.\label{eq:dual}
\end{align}

As a consequence, the effective descriptions of those structures in terms of the averaged fields should also be related to each other via the same duality  transformation, Eq.~\eqref{eq:dual}. This yields the set of electrodynamics equations
\begin{gather}
\rot{\bf H}=\frac{1}{c}\,\df{{\bf D}}{t}+\frac{4\pi}{c}\,{\bf j}\:,\mspace{8mu}\divv{\bf D}=4\pi\rho\:,\label{eq:Tchi1}\\
\rot\left({\bf E}+\tilde{\chi}\,{\bf H}\right)=-\frac{1}{c}\,\df{ }{t}\left({\bf B}-\tilde{\chi}\,{\bf D}\right)\:,\notag\\
\divv\left({\bf B}-\tilde{\chi}\,{\bf D}\right)=0\:,\label{eq:Tchi2}
\end{gather}
where ${\bf D}=\eps\,{\bf E}$, ${\bf B}=\mu\,{\bf H}$ and dual axion field reads
\begin{equation}\label{eq:Tchiexpr}
    \tilde{\chi}=-\frac{2\pi a}{\lambda}\,\int\limits_0^1\,f(sa)\,s\,ds\:,
\end{equation}
while the distribution of external sources $\rho$ and ${\bf j}$ is defined by the experimental conditions. For consistency, we provide an independent and complete derivation of Eqs.~\eqref{eq:Tchi1}-\eqref{eq:Tchiexpr} in the Supplementary Materials~\cite{supplement}. 

Crucially, the corrections appear now in the second pair of Maxwell's equations Eqs.~\eqref{eq:Tchi2} which are normally used to define the potentials. This distinguishes our model from the effective theories studied previously opening new vistas.

The effective field $\tilde{\chi}$ shares with the conventional axion field $\chi$ the same symmetry properties and, in  analogy to it, reduces to the boundary term provided $\partial\cd/\partial t=0$. Below, we explore the consequences of electrodynamics Eqs.~\eqref{eq:Tchi1}, \eqref{eq:Tchi2} and suggest experimentally observable distinctions of the axion field $\chi$ from its dual, $\tilde{\chi}$.

\section*{Key predictions}


{\bf Magnetic charges.}

Evidently, Eqs.~\eqref{eq:Tchi1}-\eqref{eq:Tchi2} describe the effective theory {\it with magnetic charge}~\cite{Shnir} when Maxwell's equations take the form
\begin{gather}
\rot{\bf H}=\frac{1}{c}\,\df{{\bf D}}{t}+\frac{4\pi}{c}\,{\bf j}\:,\mspace{8mu}\divv{\bf D}=4\pi\rho\:,\label{eq:mcharge1}\\
\rot {\bf E}=-\frac{1}{c}\,\df{{\bf B}}{t}-\frac{4\pi}{c}\,{\bf j}_m\:,\mspace{8mu}
\divv{\bf B}=4\pi\rho_m\:,\label{eq:mcharge2}
\end{gather}
where $\rho_m$ and ${\bf j}_m$ describe the density of magnetic charges and currents. In our case, those quantities are
\begin{gather}
\rho_m=\frac{1}{4\pi}\,\divv\left(\tilde{\chi}\,{\bf D}\right)\:,\label{eq:MagneticCharge}\\
{\bf j}_m=-\frac{1}{4\pi}\,\df{ }{t}\,\left(\tilde{\chi}{\bf D}\right)+\frac{c}{4\pi}\,\rot\left(\tilde{\chi}{\bf H}\right)\:,
\end{gather}
satisfying the usual continuity equation $\partial\rho_m/\partial t+\divv{\bf j}_m=0$.

The total magnetic charge of a finite volume of the medium with $\cd$ response can be readily computed as
\begin{equation}
    g\equiv\int\rho_m\,dV=\frac{1}{4\pi}\,\oint \cd\,{\bf D}\cdot{\bf n}\,df=0
\end{equation}
as $\cd$ vanishes outside of the medium. Therefore, the introduced effective description does not yield any uncompensated magnetic charges.

{\bf Reflection and transmission properties.}

Next we examine the phenomena at the boundary of a homogeneous $\cd$ medium. To that end, we integrate Eqs.~\eqref{eq:Tchi1},\eqref{eq:Tchi2} over a pillbox. Assuming no surface charges or currents, we obtain the set of boundary conditions~\cite{supplement}
\begin{gather}
    {\bf H}_{1t}={\bf H}_{2t}\:,\mspace{8mu} D_{1n}=D_{2n}\:,\label{eq:TBC1}\\
    {\bf E}_{1t}+\tilde{\chi}_1\,{\bf H}_{1t}={\bf E}_{2t}+\tilde{\chi}_2\,{\bf H}_{2t}\:,\label{eq:TBC2}\\
    B_{1n}-\tilde{\chi}_1\,D_{1n}=B_{2n}-\tilde{\chi}_2\,D_{2n}\:.\label{eq:TBC3}
\end{gather}
The tangential components of ${\bf E}$ and normal component of ${\bf B}$ now experience the discontinuity, which is not typical for the conventional media, but becomes possible here since we consider the averaged fields.

\begin{figure}[t!]
  \centering
  \includegraphics[width=0.95\linewidth]{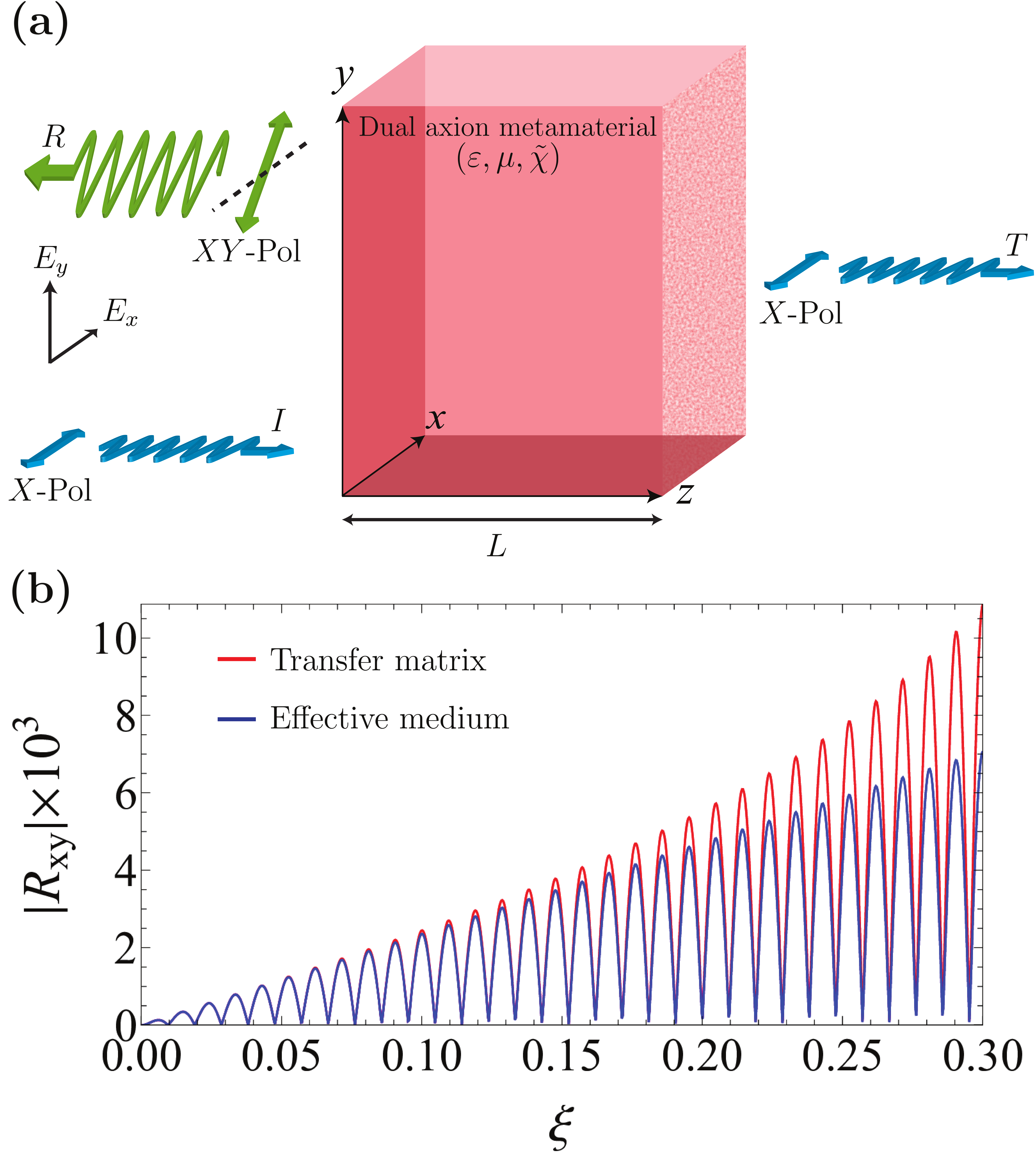}
  \caption{(a) Transmitted and reflected waves in a dual axion slab excited by the incident $x$-polarized plane wave. The polarization of reflected wave is rotated due to the Kerr effect, Eq.~(\ref{Rxy:dualaxionslab}). (b) Cross-polarized reflection coefficient for the metamaterial slab consisting of $N=50$ periods calculated via transfer matrix method (red) and effective medium approach (blue) versus $\xi=a/\lambda$. Parameters of the gyrotropic layers in the transfer matrix simulation are $\eps=1.05$, $\mu=1.05$, $f_0=0.015$. For effective medium description, we choose $\varepsilon_{\text{eff}} = 1.05$, $\mu_\text{eff} =  \mu +\frac{\pi^2}{12}|f_0|^2\xi^2$, $\tilde{\chi}_{\text{eff}} = \frac{\pi}{2} f_0 \, \xi$ and $k L = 2\pi \sqrt{\varepsilon_\text{eff} \mu_\text{eff}}  N a/\lambda$, as discussed in Supplementary Materials~\cite{supplement}.}
  \label{Fig:Dual_rxy_validation}
\end{figure}

To illustrate the physics behind the boundary conditions  Eqs.~\eqref{eq:TBC1}-\eqref{eq:TBC2}, we examine a homogeneous slab of $\cd$ medium with the thickness $L$ in vacuum calculating reflection and transmission coefficients for the linearly polarized plane wave at normal incidence as shown in Fig.~\ref{Fig:Dual_rxy_validation}(a). In close analogy to the conventional axion media, the polarization plane of reflected light is rotated as captured by the cross-polarized reflection coefficient $R_{xy}$, while cross-polarized transmission $T_{xy}$ vanishes~\cite{supplement}:
\begin{align}
     R_{xy} \hphantom{-} = - R_{yx} & =-2\cd\,\sin\left(kL\right)/\Delta,\label{Rxy:dualaxionslab}\\
     T_{xy} \hphantom{-} = \hphantom{-}T_{yx} & = 0, \label{Txy:dualaxionslab}
\end{align}
where $k$ is the propagation constant inside the slab,  
\begin{equation}
\Delta = 2iZ\,\cos kL+\left(1+Z^2+\cd^2\right)\,\sin kL
\end{equation}
and $Z=\sqrt{\mu/\eps}$ is the impedance.

To check the accuracy of the above predictions of $\cd$ electrodynamics, we simulate the metamaterial composed of pairs of oppositely magnetized gyrotropic layers using the rigorous transfer matrix method. In Fig.~\ref{Fig:Dual_rxy_validation}(b) we compare the two approaches by fixing the thickness of a metamaterial slab and varying the wavelength of the incident light such that $a/\lambda$ ratio spans the range from 0 to 0.3. We observe that the effective description of metamaterial in terms of $\cd$ is consistent provided the period of the structure $a$ is at least an order of magnitude smaller than the wavelength of light $\lambda$: $\xi = a/\lambda < 0.15$. The same applies to all remaining co- and cross-polarized reflection and transmission coefficients~\cite{supplement}.



{\bf Similarity and distinction of $\chi$ and $\cd$ responses.}

Vanishing cross-polarized transmission and nonzero cross-polarized reflection is typically considered as a smoking gun for the effective axion response~\cite{SuYangXu2023,Asadchy2024,ShuangZhang24}, which can be distinguished in this way from the magneto-optical phenomena (gyrotropy) or electromagnetic chirality. The results above suggest that $\cd$ response behaves exactly in the same way.

\begin{figure*}[ht]
    \centering
    \includegraphics[width=0.95\textwidth]{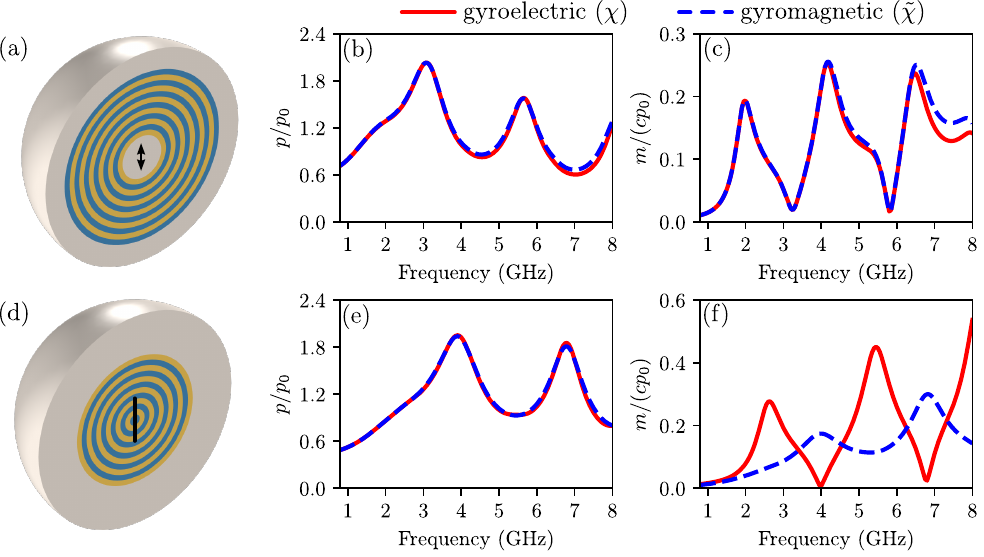}
    \caption{\label{fig:simulations} Distinguishing axion $\chi$ and dual axion $\cd$ responses of the medium via the coupling to the external sources. (a) Point electric dipole inside an air cavity within a spherical metamaterial consisting of radially magnetized layers with alternating magnetization direction. The layers possess either gyrotropic permittivity, i.e. $\mathbf{D} = \eps \mathbf{E} \pm i\,g_{\mathrm{e}} \mathbf{E} \times \hat{\mathbf{r}}$ ($\chi$ case) or gyrotropic permeability, i.e. $\mathbf{B} = \mu \mathbf{H} \pm i\,g_{\mathrm{m}} \mathbf{H} \times \hat{\mathbf{r}}$ ($\cd$ case). The thickness of each layer is $t = 2$~mm, the cavity radius is $3 t$; the material parameters $\eps = 5$, $g_{\mathrm{e}} = 4$, $\mu = 1$, $g_{\mathrm{m}} = -0.8$ ensure the mapping~\eqref{eq:mapping_in_simulations}. (d) A linear antenna of length $4 t$ inside the metamaterial. The core layer has zero  magnetization. (b,c,e,f) Effective electric and magnetic dipole moments of the system retrieved from the far field calculated for geometries (a, d), respectively. The results are normalized to the dipole moment of antenna in the absence of metamaterial.}
\end{figure*}

Quite remarkably, the parallel between $\chi$ and $\cd$ responses extends much further. As we prove~\cite{supplement}, if the electromagnetic fields are measured outside of the metamaterial, and the sources are positioned outside as well, the behavior of $\cd$ medium with permittivity $\eps$ and permeability $\mu$ is {\it indistinguishable} from the axion medium with the following effective material parameters:
\begin{gather}
 \eps_{\rm{eff}}=\frac{\eps\mu}{\mu+\eps\,\cd^2}\:,\label{eq:Mapping1}\\
 \mu_{\rm{eff}}=\mu+\eps\,\cd^2\:,\label{eq:Mapping2}\\
 \chi_{\rm{eff}}=-\frac{\eps\cd}{\mu+\eps\cd^2}\:.\label{eq:Mapping3}
\end{gather}
As elaborated in the Supplementary Materials~\cite{supplement}, the mapping Eqs.~\eqref{eq:Mapping1}-\eqref{eq:Mapping3} builds on the possibility to redefine the averaged fields ${\bf E}$ and ${\bf B}$ inside the metamaterial, since the averaged fields are the theoretical constructions and can hardly be accessed directly.




Based on this mapping, we conclude that $\chi$ and $\cd$ responses cannot be distinguished in any sort of reflection or transmission experiment typically used to capture the effective axion response.

However, despite the remarkable parallel between $\chi$ and $\cd$ fields, they are manifested differently when the external sources are introduced {\it into} the medium. Experimentally relevant examples of such situations include brehmsstrahlung, Cherenkov and transition radiation or ionization losses of charged particles moving through the material. Here, for the sake of clarity, we focus on the situation when the metamaterial sphere is excited by the external point electric dipole as depicted in Fig.~\ref{fig:simulations}(a,d). 

Theoretical analysis of this problem~\cite{supplement} suggests that the field outside of the sphere is a superposition of electric and magnetic dipole fields. Such mixing of electric and magnetic multipoles occurs due to the simultaneous breaking of $\mathcal{P}$ and $\mathcal{T}$ symmetries by the medium and hence should take place both for $\chi$ and $\cd$ metamaterials. However, the quantitative details for $\chi$ and $\cd$ media are different~\cite{supplement}.

First, we simulate the excitation of a metamaterial sphere with a spherical void, Fig.~\ref{fig:simulations}(a). If the sphere is composed of layers with gyrotropic permittivity, specifically, $\mathbf{D} = \eps \mathbf{E} \pm i\,g_{\mathrm{e}} \mathbf{E} \times \hat{\mathbf{r}}$, it features the conventional axion response $\chi$. The retrieved dipole moments for the different excitation frequencies are shown by the red lines in Fig.~\ref{fig:simulations}(b,c). On the other hand, if the layers possess gyrotropic permeability, i.e. $\mathbf{B} = \mu \mathbf{H} \pm i\,g_{\mathrm{m}} \mathbf{H} \times \hat{\mathbf{r}}$, the metamaterial sphere exhibits $\cd$ response. The calculated effective dipole moments in such case are shown by the blue lines in Fig.~\ref{fig:simulations}(b,c). To compare the two scenarios, we choose the same permittivity $\eps$ and permeability $\mu$ for both simulations, while electric and magnetic gyrotropy are linked to each other as
\begin{equation} \label{eq:mapping_in_simulations}
    g_{\mathrm{m}} = -\frac{\mu}{\eps} g_{\mathrm{e}}\,,
\end{equation}
which guarantees that the effective responses $\chi$ and $\cd$ satisfy the mapping Eqs.~\eqref{eq:Mapping1}-\eqref{eq:Mapping3} up to the second-order corrections in $\chi$ or $\cd$.

Comparing the retrieved dipole moments [Fig.~\ref{fig:simulations}(b,c)], we observe that the results for $\chi$ and $\cd$ scenarios perfectly coincide. Since the dipole is located in an air void {\it outside} of the metamaterial, this fully agrees with our theoretical analysis and provides a numerical confirmation of the mapping Eqs.~\eqref{eq:Mapping1}-\eqref{eq:Mapping3}, while small discrepancy between the two scenarios in the high-frequency region should be attributed to the limitations of the effective medium description.





Second, we analyze more interesting scenario when the external sources are introduced directly inside the metamaterial, Fig.~\ref{fig:simulations}(d). Technically, this means that any voids are smaller than the lattice period and hence the composite structure can be viewed as homogeneous. The distinction between $\chi$ and $\cd$ spheres arises already in electrostatic limit~\cite{supplement}. Specifically, $\chi$ and $\cd$ spheres excited by the point electric dipole ${\bf d}_0$ develop the effective dipole moments
\begin{gather}
    {\bf d}_{\chi}\approx \frac{3}{\eps+2}\,{\bf d}_0\:,\mspace{8mu} {\bf m}_{\chi}\approx -\frac{3\mu\chi}{\left(\eps+2\right)\left(\mu+2\right)}\,{\bf d}_0\:,\\
    {\bf d}_{\cd}\approx \frac{3}{\eps+2}\,{\bf d}_0\:,\mspace{8mu} {\bf m}_{\cd}\approx -\frac{6\cd}{\left(\eps+2\right)\left(\mu+2\right)}\,{\bf d}_0\:,
\end{gather}
where the terms proportional to $\chi^2$ or $\cd^2$ are omitted for clarity. Thus, the effective electric dipole moments coincide, while the induced magnetic dipoles differ and are not connected to each other by the mapping Eqs.~\eqref{eq:Mapping1}-\eqref{eq:Mapping3}.

To confirm that intuition, we simulate the excitation of $\chi$ and $\cd$ metamaterials in the same frequency range $f=\left(1 \div 8\right)$~GHz in the absence of voids. In this case, $\chi$ and $\cd$ span the ranges $\left(0.08 \div 0.67\right)$ and $\left(0.02 \div 0.13\right)$, respectively.
%
This time an electric dipole antenna has a finite size [Fig.~\ref{fig:simulations}(d)] such that spatial dispersion effects are suppressed and effective medium treatment remains adequate. 

In agreement with our analytical solution~\cite{supplement}, the induced electric dipole moments of both spheres coincide which  confirms that the effective medium treatment is valid in this geometry as well [Fig.~\ref{fig:simulations}(e)]. However, the induced magnetic dipole moments of $\chi$ and $\cd$ spheres are profoundly distinct featuring completely different frequency dependence [Fig.~\ref{fig:simulations}(f)] which thus distinguishes $\chi$ and $\cd$ physics.

\section*{Discussion}

These results bring us to the conclusion that there exists a distinct type of $\mathcal{P}$- and $\mathcal{T}$-odd electromagnetic response which, to the best of our knowledge, has not been identified previously. Similarly to the effective axion field, it is not manifested in the bulk of a stationary medium, but has profound boundary signatures captured by the effective theory Eqs.~\eqref{eq:Tchi1},\eqref{eq:Tchi2} with the boundary conditions Eqs.~\eqref{eq:TBC1}-\eqref{eq:TBC3}. The predicted response $\tilde{\chi}$ is a distinct physical entity as it can be  distinguished from the conventional axion field in experiments.

Here, we have put forward a single example of metamaterial realizing this physics at photonic platform. Using the same line of thought, it is also possible to design metamaterials featuring $\chi$ and $\tilde{\chi}$ responses simultaneously~\cite{supplement}. However, we anticipate that the predicted response is general and arises in many condensed matter systems, where it could have been misidentified as an effective axion field. Nonzero $\chi$ and $\tilde{\chi}$ may coexist or  continuously transform into each other upon the change of temperature or variation of the field frequency. 

Based on our analysis, we hypothesise that $\chi$ and $\cd$ fields are related to the orbital and spin contributions to the magnetization. However, quantum-mechanical theory of $\cd$ response in condensed matter remains an interesting open problem. Another open question is the possibility of quantized $\cd$ field.



\section*{Data availability}

The data that support the findings of this study are
available from the corresponding author upon reasonable request.

\begin{acknowledgments}
We acknowledge Alexey Gorlach for valuable discussions. Theoretical models were supported by the Russian Science Foundation (grant No.~23-72-10026), numerical simulations were supported by Priority 2030 Federal Academic Leadership Program. The authors acknowledge partial support by the Foundation for the Advancement of Theoretical Physics and Mathematics ``Basis''.
\end{acknowledgments}

\section*{Author contributions}

M.A.G. conceived the idea and supervised the project. T.Z.S. developed the analytical model for the dipole excitation, derived the Lagrangian description and together with E.B.-A. elaborated the mapping between $\chi$ and $\cd$ responses. E.B.-A. also analyzed the scattering from the metamaterial slab and performed transfer matrix calculations. D.A.B. performed full-wave numerical simulations of the dipole excitation. L.S. derived an effective description of $\cd$ metamaterial. M.M. contributed to transfer matrix simulation of $\chi$ and $\cd$ media. All authors contributed to the manuscript preparation.

\section*{Competing interests}

The authors declare no competing interests.

\bibliography{AxionLib}

\end{document}


\title{Supplementary Information.\\ Dual origin of effective axion response}

\author{Timur Z. Seidov}
\affiliation{School of Physics and Engineering, ITMO University, Saint  Petersburg, Russia}

\author{Eduardo Barredo-Alamilla}
\affiliation{School of Physics and Engineering, ITMO University, Saint  Petersburg, Russia}

\author{Daniel A. Bobylev}
\affiliation{School of Physics and Engineering, ITMO University, Saint  Petersburg, Russia}

\author{Leon Shaposhnikov}
\affiliation{School of Physics and Engineering, ITMO University, Saint  Petersburg, Russia}

\author{Maxim Mazanov}
\affiliation{School of Physics and Engineering, ITMO University, Saint Petersburg, Russia}

\author{Maxim A. Gorlach}
\email{m.gorlach@metalab.ifmo.ru}
\affiliation{School of Physics and Engineering, ITMO University, Saint Petersburg, Russia}


\maketitle
\widetext

\tableofcontents

\section{I. Equations of electrodynamics with $\cd$ response}

In this section, we summarize the equations of electrodynamics with $\cd$ response providing them in the three-dimensional and four-dimensional forms for future reference. We also summarize the boundary conditions which could be obtained by integrating Maxwell's equations over a pillbox. The detailed discussion and derivation of those equations is provided in the main text and the sections below.

\begin{center}
{\bf Maxwell's equations with both $\chi$ and $\tilde\chi$ in a vector form}
\end{center}

\begin{gather}
\rot\left( {\bf H} -\chi\,{\bf E}\right)=\frac{1}{c}\,\df{ }{t}\,\left({\bf D}+\chi\,{\bf B}\right)+\frac{4\pi}{c}\,{\bf j}\:, \label{eq:ChiTchi1}\\
\divv\left({\bf D}+\chi\,{\bf B}\right)=4\pi\rho\:, \label{eq:ChiTchi2} \\
\rot\left({\bf E}+\tilde{\chi}\,{\bf H}\right)=-\frac{1}{c}\,\df{ }{t}\left({\bf B}-\tilde{\chi}\,{\bf D}\right)\:,\label{eq:ChiTchi3}\\
\divv\left({\bf B}-\tilde{\chi}\,{\bf D}\right)=0\:. \label{eq:ChiTchi4}
\end{gather}
Here, ${\bf D}=\eps\,{\bf E}$ and ${\bf B}=\mu\,{\bf H}$. $\rho$ and ${\bf j}$ are the external charge and current densities. $\chi$ and $\cd$ quantify the effective axion and dual axion fields. In general, both could depend on time and coordinates. 

In the absence of sources Eqs.~\eqref{eq:ChiTchi1}-\eqref{eq:ChiTchi4} are invariant under the following electromagnetic duality transformation
\begin{gather}
    {\bf E} \longrightarrow {\bf H}\:, \quad {\bf H} \longrightarrow -{\bf E}\:, \label{eq:Duality1}\\
    {\bf D} \longrightarrow {\bf B}\:, \quad {\bf B} \longrightarrow -{\bf D}\:, \label{eq:Duality2}\\
    \varepsilon \longleftrightarrow \mu \:, \quad \chi \longleftrightarrow \tilde\chi\:.\label{eq:Duality3}
\end{gather}
Note that this transformation appears here in a different context compared to the Tellegen media literature \cite{Lindell_duality,Prudencio_duality,Tretyakov_duality}, where it is used as a computational tool to solve for electromagnetic wave propagation in Tellegen media.


\begin{center}
{\bf Boundary conditions in the medium with both $\chi$ and $\tilde\chi$}
\end{center}

By integrating Eqs.~\eqref{eq:ChiTchi1}-\eqref{eq:ChiTchi4} over a pillbox and assuming that the external sources are absent at the boundaries, we recover the following set of boundary conditions:
\begin{gather}
    {\bf H }_{1t} - \chi_1 \, {\bf E }_{1t}={\bf H }_{2t} - \chi_2 \, {\bf E }_{2t}\:, \label{eq:BC1}\\
    D_{1n}+ \chi_1 \, B_{1n} = D_{2n}+ \chi_2 \, B_{2n} \:,\label{eq:BC2}\\
    {\bf E}_{1t}+\tilde{\chi}_1\,{\bf H}_{1t}={\bf E}_{2t}+\tilde{\chi}_2\,{\bf H}_{2t}\:,\label{eq:TBC1}\\
    B_{1n}-\tilde{\chi}_1\,D_{1n}=B_{2n}-\tilde{\chi}_2\,D_{2n}\:.\label{eq:TBC2}
\end{gather}
The subscripts $t$ and $n$ indicate the components of the fields tangential and normal to the boundary, respectively.

\begin{center}
    {\bf Maxwell's equations with both $\chi$ and $\tilde\chi$ in the covariant form}
\end{center}

The same equations can be written in the 4D form. Here, we utilize the conventions $\mu,\nu,\alpha,\beta,\dots=0,1,2,3$; $i,j,k,l,\dots=1,2,3$; $g_{\mu\nu}=\text{diag}\left(1,-1,-1,-1\right)$ as in the classical textbook Ref.~\cite{Landau}.

Four-vectors are defined as $x^\mu=\left(ct,{\bf r}\right)$, $j^\mu=\left(c\rho,{\bf j}\right)$, $A^\mu=\left(\varphi,{\bf A}\right)$. The respective four-derivative is $\partial_\mu=\left(\frac{1}{c}\,\df{ }{t},{\bf \nabla}\right)$. Electromagnetic field tensor is defined by $F_{\mu\nu}=\partial_{\mu}A_\nu-\partial_\nu A_{\mu}$. Fully antisymmetric fourth rank tensor $e^{\mu\nu\alpha\beta}$ is defined by the condition $e^{0123}=1$. Accordingly, $e_{0123}=-1$. We also use the expression for the convolution of two Levi-Civita symbols~\cite{Landau}
\begin{equation}
e^{\mu\nu\alpha\beta}\,e_{\sigma\rho\alpha\beta}=-2\,\left(\delta^{\mu}_{\sigma}\,\delta^{\nu}_{\rho}-\delta^{\mu}_{\rho}\,\delta^{\nu}_{\sigma}\right)\:.
\end{equation}
Dual electromagnetic field tensor reads: $\tilde{F}^{\mu\nu}=\frac{1}{2}\,e^{\mu\nu\alpha\beta}\,F_{\alpha\beta}$. This expression can be inverted to yield $F^{\mu\nu}=-\frac{1}{2}\,e^{\mu\nu\alpha\beta}\,\tilde{F}_{\alpha\beta}$.

In the three-dimensional space, we also define a second rank antisymmetric tensor dual to the given vector ${\bf B}$: $B^{\times}_{ij}=e_{ijk}\,B^k$, where $e^{ijk}$ is Levi-Chivita symbol with the normalization $e^{123}=1$, $e_{123}=-1$. Hence,
\begin{equation}
    B^{\times}=
    \begin{pmatrix}
        0 & -B_z & B_y\\
        B_z & 0 & -B_x\\
        -B_y & B_x & 0
    \end{pmatrix}\:,
\end{equation}
where $B_x$, $B_y$ and $B_z$ denote the contravariant components of the ${\bf B}$ vector $B^1$, $B^2$ and $B^3$. As a result, the action of $B^{\times}$ on the arbitrary vector ${\bf a}$ is equivalent to the vector product ${\bf B}\times{\bf a}$.

In such case, the explicit expressions for electromagnetic field tensor and its dual take the following form:
\begin{gather}
    F_{\mu\nu}=\begin{pmatrix}
        0 & {\bf E}\\
        -{\bf E} & B^\times
    \end{pmatrix}\:,
    \mspace{10mu}
    F^{\mu\nu}=\begin{pmatrix}
        0 & -{\bf E}\\
        {\bf E} & B^{\times}
    \end{pmatrix}\:\\
    \tilde{F}_{\mu\nu}=\begin{pmatrix}
        0 & {\bf B}\\
        -{\bf B} & -E^\times
        \end{pmatrix}\:,
    \mspace{10mu}
    \tilde{F}^{\mu\nu}=\begin{pmatrix}
        0 & -{\bf B}\\
        {\bf B} & -E^\times
    \end{pmatrix}\:.
\end{gather}


Since we consider the fields in the continuous medium, it is convenient to introduce one more four-tensor composed of the components of ${\bf D}$ and ${\bf H}$ arranged in a similar way:
\begin{gather}
    H_{\mu\nu}=\begin{pmatrix}
        0 & {\bf D}\\
        -{\bf D} & H^\times
    \end{pmatrix}\:,
    \mspace{10mu}
    H^{\mu\nu}=\begin{pmatrix}
        0 & -{\bf D}\\
        {\bf D} & H^{\times}
    \end{pmatrix}\:\\
    \tilde{H}_{\mu\nu}=\begin{pmatrix}
        0 & {\bf H}\\
        -{\bf H} & -D^\times\:,
        \end{pmatrix}
    \mspace{10mu}
    \tilde{H}^{\mu\nu}=\begin{pmatrix}
        0 & -{\bf H}\\
        {\bf H} & -D^\times
    \end{pmatrix}\:.
\end{gather}
%
%
In turn, the connection between $H$ and $F$ tensors is derived from the constitutive relations. As we assume here ${\bf D}=\eps\,{\bf E}$ and ${\bf B}=\mu\,{\bf H}$, we recover
\begin{equation} H_{\mu\nu}=d_{\mu\alpha}\,d_{\nu\beta}\,F^{\alpha\beta}\:,
\end{equation}
where
\begin{equation}
    d_{\mu\nu}=\frac{1}{\sqrt{\mu}}\,
    \begin{pmatrix}
     \eps\,\mu & 0\\
     0 & -I
    \end{pmatrix}\:,
\end{equation}
and $I$ is a $3\times 3$ identity matrix. We note that the duality transformation Eqs.~\eqref{eq:Duality1}-\eqref{eq:Duality3} connects the introduced tensors in the following way
\begin{gather}
   F^{\mu\nu}\rightarrow \tilde{H}^{\mu\nu},\mspace{6mu} \tilde{H}^{\mu\nu}\rightarrow -F^{\mu\nu}\:,\\
   H^{\mu\nu}\rightarrow\tilde{F}^{\mu\nu},\mspace{6mu} \tilde{F}^{\mu\nu}\rightarrow -H^{\mu\nu}\:.
\end{gather}


Using the above designations, the equations governing electrodynamics of the medium with $\chi$ and $\cd$ responses, Eqs.~\eqref{eq:ChiTchi1}-\eqref{eq:ChiTchi4} are recast in the form
\begin{gather}
    \partial_\mu (H^{\mu\nu}+\chi \, \tilde F^{\mu\nu})=\frac{4\pi}{c}\,j^{\nu}\:, \label{eq:Covariant1} \\
    \partial_\mu(\tilde F^{\mu\nu}-\cd \, H^{\mu\nu})=0\:. \label{eq:Covariant2}
\end{gather}
Examining this formulation, we identify $\cd$ term as one restoring the dual symmetry between the two pairs of Maxwell's equations in the presence of the axion field.

\section{II. Metamaterial with $\cd$ response}\label{app: cd metamat}


This section supplements the discussion in the main text  and provides a complete and rigorous derivation of $\cd$ response in a specific metamaterial. We examine a multilayered structure composed of gyrotropic layers with the scalar and constant permittivity $\eps$ and permeability of the form
\begin{equation}\label{eq:materialEq}
\hat{\mu}(z)=\begin{pmatrix}
\mu & i\,f(z) & 0\\
-i\,f(z) & \mu & 0\\
0 & 0 & \mu
\end{pmatrix}.
\end{equation}
Here, the gyrotropy $f(z)$ is a periodic function with the period $a$, so that its Fourier expansion
\begin{equation}
    f(z) =  \sum_{n=-\infty}^{+\infty} f_n\,e^{i n b z}, \quad f_n =\frac{1}{a} \int_0^a f(z)\,e^{-i n b z}\,dz\:, 
\end{equation}
where $b=2\pi/a$ is the reciprocal lattice period, and $f_0 \equiv 0$ as we consider magnetization distribution with the zero average. For instance, such situation occurs in antiferromagnets. Having these expressions, we expand the permeability in the Fourier series:
\begin{equation}\label{eq:cdhatmu}
\hat{\mu}(z) = \sum_{n=-\infty}^{+\infty}\hat{\mu}_n\; e^{i n b z}, \quad
\hat{\mu}_n = \mu\;\hat{I} \; \delta_{n0} - i f_n \; \bas{z}^{\times}, \quad 
\bas{z}^{\times} =\begin{pmatrix}
0 & -1 & 0\\
1 & 0& 0\\
0 & 0 & 0
\end{pmatrix}
\end{equation}
where tensor $\bas{z}^{\times}$ acts as a cross-product with $\bas{z}$. Using the periodicity of the structure, we present the field inside the metamaterial in the following form
\begin{equation}\label{eq:Floquetexp}
    \begin{pmatrix}
    {\bf E}({\bf r}, t)\\
    {\bf H}({\bf r}, t)\\
    {\bf D}({\bf r}, t)\\
    {\bf B}({\bf r}, t)
    \end{pmatrix}
    =
    \sum\limits_{n=-\infty}^{\infty}\,
    \begin{pmatrix}
    {\bf E}_n({\bf r})\\
    {\bf H}_n({\bf r})\\
    {\bf D}_n({\bf r})\\
    {\bf B}_n({\bf r})    
    \end{pmatrix}
    \,e^{i{\bf k}^{(n)}\cdot{\bf r} - i \omega t}\:,
\end{equation}
where $\vc{k}^{(n)}=\skob{k_x, 0, k_z + n b}$. $k_z$ and $k_x$ are the components of the Bloch vector parallel and perpendicular to the layers, respectively. The component $k_y$ can always be set to zero by the choice of the coordinate system due to the rotational symmetry of the system with respect to the $z$ axis.

The components $\vc{E}_0,\;\vc{H}_0,\;\vc{D}_0$ and $\vc{B}_0$  are averaged (macroscopic) fields, while amplitudes with non-zero $n$ describe rapidly oscillating Floquet harmonics. We consider the metamaterial limit when the structure period is much less than the wavelength. In the other words, $\xi=a/\lambda=q/b$ ($q=\om/c$) plays the role of small parameter.

%
We aim to derive the macroscopic description of the structure in terms of the averaged fields. To that end, we need to compute the amplitudes of higher-order Floquet harmonics expanding them in powers of small parameter $\xi=a/\lambda=q/b$.

Microscopic fields in the structure satisfy the conventional set of equations
\begin{gather}
\nabla\left(\divv{\bf H}\right)-\Delta {\bf H}=q^2\,\eps\,{\bf B},\label{eq:waveEq}\\
\text{{div\rm}} \,\vc{B} = 0, \label{eq:divB}\\
\vc{B}=\hat{\mu}\,\vc{H}\label{eq:Mu}
\end{gather}
with $q=\omega/c$.  Substituting expansion \eqref{eq:Floquetexp} into Eqs.\eqref{eq:waveEq},\eqref{eq:divB},\eqref{eq:Mu} above, we obtain the set of scalar equations for the respective Floquet harmonics
\begin{gather}
    \left(k_z^{(n)}\right)^2\,H_{nx}-k_x k_z^{(n)}\,H_{nz}=q^2\,\eps\,B_{nx}\:,\label{eq:Aux1}\\
    \left[k_x^2+\left(k_z^{(n)}\right)^2\right]\,H_{ny}=q^2\,\eps\,B_{ny}\:,\label{eq:Aux2}\\
    -k_x\,k_z^{(n)}\,H_{nx}+k_x^2\,H_{nz}=q^2\,\eps\,B_{nz}\:,\label{eq:Aux3}\\
    k_x\,B_{nx}+k_z^{(n)}\,B_{nz}=0\:,\label{eq:Aux4}\\
    B_{nx}=\mu\,H_{nx}+i\,\sum\limits_{n'\not=n}\,f_{n-n'}\,H_{n'y}\:,\label{eq:Aux5}\\
    B_{ny}=\mu\,H_{ny}-i\,\sum\limits_{n'\not=n}\,f_{n-n'}\,H_{n'x}\:,\label{eq:Aux6}  \\
    B_{nz}=\mu\,H_{nz}\:.\label{eq:Aux7}    
\end{gather}
with $k_z^{(n)} = k_z + n b$. From this system of equations \eqref{eq:Aux1}-\eqref{eq:Aux7}, we express $H_{nx}, H_{ny}$ and $H_{nz}$  up to second order of the small parameter $\xi$.

Equation~\eqref{eq:Aux2} yields $H_{ny}\approx q^2 \eps/(n^2\,b^2)\,B_{ny}+O(\xi^3)$. From Eq.~\eqref{eq:Aux6}, we recover $B_{ny}=-if_n\,H_{0x}+O(\xi^2)$. Combining these two results, we obtain
\begin{equation}\label{eq:HnyFl}
    H_{ny}=-if_n\,\frac{\eps\,q^2}{n^2\,b^2}\,H_{0x}+O(\xi^3)\:.
\end{equation}

Next, Eq.~\eqref{eq:Aux5} yields $B_{nx}=if_n\,H_{0y}+O(\xi^2)$ and Eq.~\eqref{eq:Aux4} allows to calculate $B_{nz}$: 
\begin{equation}\label{eq:BnzFl}
 B_{nz}=-\frac{k_x}{k_z^{(n)}}\,B_{nx}=-if_n\,\frac{k_x}{nb}\,\left(1-\frac{k_z}{nb}\right)\,H_{0y}+O(\xi^3). 
\end{equation}
Using Eq.~\eqref{eq:Aux7}, we immediately evaluate
\begin{equation}\label{eq:HnzFl}
    H_{nz}=-if_n\,\frac{k_x}{\mu\,nb}\,\left(1-\frac{k_z}{nb}\right)\,H_{0y}+O(\xi^3)\:.
\end{equation}
Finally, we use Eq.~\eqref{eq:Aux1} to calculate $H_{nx}$ via already found $B_{nx}$ and $H_{nz}$ which yields
\begin{equation}\label{eq:HnxFl}
    H_{nx}=\frac{if_n}{n^2\,b^2}\,\left(\eps\,q^2-\frac{k_x^2}{\mu}\right)\,H_{0y}+O(\xi^3)\:.
\end{equation}

Since $\text{rot}\,\vc{H}=-iq\,\vc{D}$, we can calculate higher-order Floquet harmonics of $\vc{D}$ and $\vc{E}$: $\vc{D}_n=-\left[\vc{k}^{(n)}\times\vc{H}_n\right]/q$, $\vc{E}=\vc{D}/\eps$. This yields:
\begin{align}
    &E_{nx}=-if_n\,\frac{q}{nb}\,H_{0x}+O(\xi^2)\:,\label{eq:DnxFl}\\
    &E_{ny}=-if_n\,\frac{q}{nb}\,H_{0y}+O(\xi^2)\:,\label{eq:DnyFl}\\
    &D_{nz}=if_n\,\frac{k_x\,q\,\eps}{n^2\,b^2}\,H_{0x}+O(\xi^3)\:,\label{eq:DnzFl}\\
    &D_{0z}=-\frac{k_x\,H_{0y}}{q}\:.\label{eq:D0z}
\end{align}

Having the explicit expressions for the Floquet harmonics, we derive now the boundary conditions for the averaged fields. The microscopic fields in the structure satisfy the conventional boundary conditions:
\begin{align}
&\left.{\bf H}_{t}\right|_{z=0}={\bf H}_t^{{\rm out}}\:,
&\left.D_{z}\right|_{z=0}=D_z^{{\rm out}}\:,\label{eq:MCond1}\\
&\left.{\bf E}_{t}\right|_{z=0}={\bf E}_t^{{\rm out}}\:
&\left.B_{z}\right|_{z=0}=B_z^{\rm out}\:.\label{eq:MCond2}
\end{align}

First we examine the boundary conditions for the tangential component of the macroscopic $\vc{H}$ field and normal component of the macroscopic $\vc{D}$. We substitute expressions for Floquet harmonics of the fields Eqs.\eqref{eq:HnxFl},\eqref{eq:HnyFl},\eqref{eq:DnzFl} into Eq.\eqref{eq:MCond1} and obtain
\begin{align}
    & \vc{H}_{0t} + O(\xi^2) = {\bf H}_t^{{\rm out}}, & 
    D_{0z}  + O(\xi^2) = D_z^{{\rm out}}.
\end{align}
Hence, these macroscopic fields are continuous up to second order of small parameter $\xi$. However, the second pair of boundary conditions Eq.~\eqref{eq:MCond2} is modified, as some of the Floquet harmonics are proportional to the first power of small parameter. Using Eqs.\eqref{eq:BnzFl},\eqref{eq:DnxFl},\eqref{eq:DnyFl},\eqref{eq:D0z}, we derive
\begin{align}\label{eq:metamatBC}
    & \vc{E}_{0t} + \cd \vc{H}_{0t} + O(\xi^2) = {\bf E}_t^{{\rm out}}, & 
    B_{0z} - \cd D_{0z} + O(\xi^2) = B_z^{{\rm out}}.
\end{align}
where
\begin{equation}\label{eq:cd}
    \cd = - i \frac{q}{b} \sum_{n \neq 0} \frac{f_n}{n}
\end{equation}

At this point, we stress that the derived boundary conditions are the boundary conditions for $\cd$ electrodynamics, Eqs.~\eqref{eq:BC1}-\eqref{eq:TBC2} with zero axion response $\chi$. Hence, up to the first order in $\xi$ the designed metamaterial can be viewed as $\cd$ medium. Therefore, it should exhibit all the characteristic phenomena including Kerr rotation for the reflected plane waves. We verify this by numerical simulations in the main text and the sections below.


\section{III. Metamaterial with both $\chi$ and $\cd$}

In this section, we discuss the possibility to achieve $\chi$ and $\cd$ responses in a metamaterial simultaneously. As we have seen before, the gyrotropic permittivity $\hat{\eps}$ of the layers is responsible for $\chi$ response, while $\hat{\mu}$ enables $\cd$ field. However, there exist natural materials with both gyrotropic $\hat{\eps}$ and $\hat{\mu}$. We therefore anticipate that the metamaterial composed of such bi-gyrotropic layers will exhibit $\chi$ and $\cd$ responses simultaneously.

Accordingly, we examine a metamaterial whose permittivity and permeability are spatially modulated:
\begin{equation}
\hat{\eps}(z)=\begin{pmatrix}
\eps & i\,g(z) & 0\\
-i\,g(z) & \eps & 0\\
0 & 0 & \eps
\end{pmatrix},\quad
\hat{\mu}(z)=\begin{pmatrix}
\mu & i\,f(z) & 0\\
-i\,f(z) & \mu & 0\\
0 & 0 & \mu
\end{pmatrix}.
\end{equation}

We assume that $f(z)$ and $g(z)$ are periodic functions of the same period $a$. In that case, we may still seek a solution in the form of the Floquet series.

For clarity, we focus on the normal incidence scenario $\vc{k} = (0, 0, k_z)$ when all the fields are in $Oxy$ plane and apply the same Floquet expansion for the electromagnetic fields \eqref{eq:Floquetexp}. Using Ampere's and Faraday's laws and keeping the terms up to second order in small parameter $\xi=a/\lambda\ll1$, we derive 
\begin{align}\label{eq: curl eq}
    &\vc{E}_n=-\frac{q}{nb}\left(1-\frac{k_z}{nb}\right)\,\left[\bas{z}\times\vc{B}_n\right]+O(\xi^3),\nonumber\\
    &\vc{H}_n=\frac{q}{nb}\left(1-\frac{k_z}{nb}\right)\,\left[\bas{z}\times\vc{D}_n\right]+O(\xi^3).
\end{align}

From the constitutive relations we can express $n$-th Floquet harmonics of $\vc{B},\;\vc{D}$ fields
\begin{align}\label{eq: const eq}
&\vc{B}_n=\hat{\mu}_n\vc{H}_0+\sum_{n'\neq n}\hat{\mu}_{n'}\vc{H}_{n-n'},\quad \hat{\mu}_n=-if_n\,\bas{z}^{\times}  \nonumber\\
&\vc{D}_n=\hat{\eps}_n\vc{E}_0+\sum_{n'\neq n}\hat{\eps}_{n'}\vc{E}_{n-n'},\quad \hat{\eps}_{n}=-ig_n\,\bas{z}^{\times}
\end{align}


Combining Eqs.\eqref{eq: curl eq},\eqref{eq: const eq}, we obtain expressions for the Floquet harmonics of the fields
\begin{align}
    & \vc{E}_n = -if_n\,\frac{q}{nb}\vc{H}_0 + O(\xi^2) \label{eq:Etfl}\:,\\
    & \vc{H}_n = ig_n\,\frac{q}{nb} \vc{E}_0 + O(\xi^2) \label{eq:Htfl}\:.
\end{align}
%
In the case of normal incidence we are interested only in the boundary conditions for the  tangential components of the electromagnetic field, while the conditions for the normal components are fulfilled automatically. 
Substituting Eqs.\eqref{eq:Etfl},\eqref{eq:Htfl} into conventional boundary conditions for microscopic fields Eqs.\eqref{eq:MCond1},\eqref{eq:MCond2} and keeping terms up to first order of small parameter $\xi$ we derive 
\begin{align}
   & \vc{E}^{\text{out}}=\vc{E}_0+\skob{-i\frac{q}{b}\sum_{n\neq0}\frac{f_n}{n}}\vc{H}_0,\\
   & \vc{H}^{\text{out}}=\vc{H}_0-\skob{-i\frac{q}{b}\sum_{n\neq0}\frac{g_n}{n}}\vc{E}_0.
\end{align}

The boundary conditions above coincide with the boundary conditions for the media with both $\chi$ and $\cd$ Eq.~\eqref{eq:BC1}, \eqref{eq:TBC1}, where 
\begin{gather}
    \chi=-i\frac{q}{b}\sum_{n\neq0}\frac{g_n}{n}\:,\\
    \cd=-i\frac{q}{b}\sum_{n\neq0}\frac{f_n}{n}\:.
\end{gather}


\section{IV. Mapping between $\c$ and $\cd$}\label{app:mapping}

As discussed in the main text, there exists a remarkable parallel between the electromagnetic properties of the medium with $\cd$ response, permittivity $\eps$ and permeability $\mu$ and a medium with renormalized permittivity $\eps'$, permeability $\mu'$ and an axion response $\chi'$. We derive the detailed mapping between the two descriptions below. For the sake of generality, we assume that all of the material parameters depend on time and coordinates.

This derivation relies on the following principle: the averaged electromagnetic fields arising in the effective description of metamaterials are auxiliary quantities and cannot be measured directly inside the structure. The only quantity of physical relevance is the combination of electric and magnetic fields entering the Lorentz force and defining the motion of external charges. Another crucial requirement to the transformation is to keep the fields outside of the metamaterial intact, as those fields are measured experimentally.

In this spirit, given the fields ${\bf E}$, ${\bf B}$, ${\bf D}$ and ${\bf H}$ in $\cd$ medium, we introduce the redefined fields as
\begin{gather}
  \vc{E}^\prime=\vc E + \cd\,\vc H , \label{eq:E_eff} \\
  \vc{B}^\prime=\vc B - \cd \vc D \label{eq:B_eff},
\end{gather}
where $\vc{D}=\eps\vc{E}$ and $\vc{B}=\mu\vc{H}$. The equations governing the electrodynamics of $\cd$ medium read:
\begin{gather}
\rot{\bf H}=\frac{1}{c}\,\df{{\bf D}}{t}+\frac{4\pi}{c}\,{\bf j}\:,\mspace{8mu}\divv{\bf D}=4\pi\rho\:,\label{eq:MaxwellTchi1}\\
\rot\left({\bf E}+\tilde{\chi}\,{\bf H}\right)=-\frac{1}{c}\,\df{ }{t}\left({\bf B}-\tilde{\chi}\,{\bf D}\right)\:,\notag\\
\divv\left({\bf B}-\tilde{\chi}\,{\bf D}\right)=0\:.\label{eq:MaxwellTchi2}
\end{gather}

Clearly, the transformation Eqs.~\eqref{eq:E_eff}-\eqref{eq:B_eff} converts the second pair of Maxwell's equations Eq.~\eqref{eq:MaxwellTchi2} to the standard form
\begin{gather}
\rot\vc{E}'=-\frac{1}{c}\,\df{\vc{B}'}{t}\:,\mspace{6mu}
\divv\vc{B}'=0\:.\label{eq:MaxwellChi1}
\end{gather}
Next we examine the transformation of the first pair. To this end we introduce the redefined effective parameters
\begin{gather}
    \varepsilon'= \frac{\varepsilon \mu }{\varepsilon \cd^2+\mu } ,  \label{eq:eps_eff} \\
    \mu'= \varepsilon \cd^2+\mu , \label{eq:mu_eff} \\
    \chi'=-\frac{\varepsilon  \cd }{\varepsilon \cd^2+\mu} \label{eq:chi_eff}
\end{gather}
and calculate $\vc{H}$ and $\vc{D}$ in terms of the redefined fields Eqs.~\eqref{eq:E_eff},\eqref{eq:B_eff} and material parameters Eqs.~\eqref{eq:eps_eff}-\eqref{eq:chi_eff}. Straightforward calculation yields:
\begin{gather}
 \vc{H}=-\chi'\,\vc{E}'+\vc{H}'\:,\label{eq:H_subs_EB}\\\vc{D}=\vc{D}'+\chi'\,\vc{B}'\:.\label{eq:D_subs_EB}
\end{gather}

This immediately transforms the first pair of Maxwell's equations into
\begin{gather}
    \rot \left( \vc{H}' - \chi' \vc{E}' \right)=\frac{1}{c}\,\df{{}}{t}\left(\vc{D}'+\chi'\vc{B}'\right)+\frac{4\pi}{c}\,{\bf j}\:, \\
    \divv \left(\vc{D}'+\chi'\vc{B}'\right)=4\pi \rho \:,
\end{gather}
where $\vc{D}'=\varepsilon' \vc{E}'$ and $\vc{B}'=\mu'\vc{H}'$. These equations coincide with a pair of equations~\eqref{eq:ChiTchi1}-\eqref{eq:ChiTchi2} of the usual axion electrodynamics. Taken together with \eqref{eq:MaxwellChi1}, this proves the equivalence between dual axion response $\cd$ and an axion response $\chi'$. Hence, we conclude that \textit{in the absence of sources the dual axion medium is equivalent to the conventional axion medium} and any experiment measuring the fields outside of the metamaterial cannot distinguish between the dual axion medium with the material parameters $(\eps,\mu,\cd)$ and an  axion medium with parameters $(\eps',\mu',\c')$.

However, the parallel between the two responses breaks down once the sources are introduced into the medium. As the Lorentz force is an observable quantity, the redefinition of the fields $\vc{E}$ and $\vc{B}$ should necessarily be accompanied by the redefinition of the external charge and current densities to ensure that the Lorentz force remains invariant.

Since the mapping transforms electric fields into a combination of electric and magnetic fields, we assume that the electric charges are mapped onto a combination of electric and {\it magnetic} sources. More specifically,
\begin{align*}
    \begin{cases}
       \text{Sources:}\: \rho_{e}, \vc j_{e} \\
       \text{Fields:}\: \vc E, \vc B, \vc D, \vc H \\
       \text{Parameters:}\: \eps,\mu,\cd
    \end{cases} &
    \quad \quad \longrightarrow &
    \begin{cases}
       \text{Sources:}\: \rho_{e}', \vc j_{e}',\rho_{m}', \vc j_{m}' \\
       \text{Fields:}\: \vc E', \vc B', \vc D', \vc H' \\
       \text{Parameters:}\: \eps_{\mathrm{eff}},\mu_{\mathrm{eff}},\c_{\mathrm{eff}}
    \end{cases} \\
    \vc f=\rho_{e}\vc E+\frac{1}{c}[\vc j_{e}\times \vc B] & \quad \quad \longrightarrow & \vc f'=\rho_{e}'\vc E'+\frac{1}{c}[\vc j_{e}'\times \vc B'] + \rho_{m}'\vc H' - \frac{1}{c}[\vc j_{m}'\times \vc D']
\end{align*}
We require that the equivalence $\vc f= \vc f'$ holds for any velocity of the charges and therefore
\begin{gather}
    \rho_{e} \vc E= \rho_{e}'\vc E' + \rho_{m}'\vc H' = \rho_{e}'(\vc E + \cd\vc H) +  \rho_{m}' {\mu'}^{-1}(\mu \vc H - \cd \eps \vc E)\:, \\
    \vc j_{e} \times \vc B = \vc j_{e}'\times \vc B' + \vc j_{m}'\times \vc D' =  \vc j_{e}'\times (\vc B - \cd \vc D) - \vc j_{m}'\times \eps'(\eps^{-1} \vc D + \cd \mu^{-1} \vc B)\:.
\end{gather}
Since electric and magnetic components are independent, we derive a system
\begin{equation}
    \begin{cases}
        \rho_{e}=\rho_{e}'+\c'\rho_{m}' \\
        \cd \rho_{e}' + \mu'^{-1}\mu \: \rho_{m}'= 0 \\
        \vc j_{e} = \vc j_{e}' + \c'\vc j_{m}' \\
        \cd \vc j_{e}' + \mu'^{-1}\mu \: \vc j_{m}'= 0\:,
    \end{cases}
\end{equation}
the solution of which reads
\begin{equation}
    (\rho_{m}', \vc j_{m}')= -\cd(\rho_{e},\vc j_{e}), \quad (\rho_{e}', \vc j_{e}')=\frac{\mu}{\mu'}(\rho_{e},\vc j_{e}). \label{eq:charge_map}
\end{equation}
Note that the factor $\mu/\mu'$ can also be presented as $(1+\chi'\cd)$.

This calculation suggests that in order to preserve physically observable Lorenz force acting on the external sources, we need to transform the source density simultaneously with the fields and the material parameters. As a consequence, the equations governing the dynamics of the redefined fields read
\begin{gather}
    \rot \left( \vc{H}' - \chi' \vc{E}' \right)=\frac{1}{c}\,\df{{}}{t}\left(\vc{D}'+\chi'\vc{B}'\right)+\frac{4\pi}{c}\,\vc j_{e}'\:, \\
    \divv \left(\vc{D}'+\chi'\vc{B}'\right)=4\pi \rho_{e}' \:, \\
    \rot\vc{E}'=-\frac{1}{c}\,\df{\vc{B}'}{t}\:-\frac{4\pi}{c}\vc j_{m}',\\
    \divv\vc{B}'=4\pi\rho_{m}'\:.
\end{gather}

We conclude, that in presence of the external sources~-- electric charges and currents~-- the dual axion medium can be effectively described as an ordinary axion medium, but  with the external electric and magnetic charges. Similar conclusion could be made from a purely topological perspective - since field mapping \eqref{eq:E_eff},\eqref{eq:B_eff} effectively maps potential field onto the solenoidal field, one needs to introduce effective magnetic charges to fix that. 


We can make even more general statement: in the absence of sources, a given medium with spacetime-dependent $\c$ and $\cd$ responses can be described with a single effective parameter $\c'$. To prove that, we apply the same redefinition of the fields Eqs.~\eqref{eq:E_eff},\eqref{eq:B_eff} to the material with both types of responses. The starting equations are
\begin{gather}
    \rot (\vc H - \c \vc E)=\frac{1}{c}\frac{\partial}{\partial t}(\vc D + \c \vc B),\mspace{10mu} \divv\left({\bf D}+\chi\,\vc{B}\right)=0\:, \label{eq:Maxwell_chi_tchi_1}\\
    \rot (\vc E + \cd \vc H)=-\frac{1}{c}\frac{\partial}{\partial t}(\vc B - \cd \vc D),\mspace{10mu} \divv\left({\bf B}-\cd\,\vc{D}\right)=0\:. \label{eq:Maxwell_chi_tchi_2}
\end{gather}
Next we identify the effective material parameters as
\begin{gather}
    \mu'=\mu \frac{1+\frac{\varepsilon}{\mu}\cd^2}{1+\c \cd}, \label{eq:MappingG1}\\
    \varepsilon'=\varepsilon \frac{1+\c\cd}{1+\frac{\varepsilon}{\mu}\cd^2}, \label{eq:MappingG2}\\
    \c'=\frac{\c - \frac{\varepsilon}{\mu}\cd}{1+\frac{\varepsilon}{\mu}\cd^2}\label{eq:MappingG3}
\end{gather}
and check the behavior of the two field combinations:
\begin{gather}
    \vc H' - \c' \vc E' = \frac{1+\c\cd}{1+\frac{\varepsilon}{\mu}\cd^2}\mu^{-1}(\vc B-\varepsilon \cd \vc E)-\frac{\c - \frac{\varepsilon}{\mu}\cd}{1+\frac{\varepsilon}{\mu}\cd^2}(\vc E + \frac{\cd}{\mu}\vc B)=\vc H-\c \vc E,\\
   \vc D'+\c' \vc B'=\varepsilon \frac{1+\c\cd}{1+\frac{\varepsilon}{\mu}\cd^2}(\vc E + \frac{\cd}{\mu}\vc B)+\frac{\c - \frac{\varepsilon}{\mu}\cd}{1+\frac{\varepsilon}{\mu}\cd^2}(\vc B-\varepsilon \cd \vc E)=\vc D+\c \vc B.
\end{gather}
From this we observe that the first pair of equations~\eqref{eq:Maxwell_chi_tchi_1} remains invariant, where all fields and material parameters are replaced by their redefined values. At the same time we see that the equation \eqref{eq:Maxwell_chi_tchi_2} is precisely \eqref{eq:MaxwellChi1}. Hence, in the absence of the sources our procedure allows to describe the medium with $\chi$ and $\cd$ responses with a single nonreciprocal magneto-electric parameter.

\section{V. Lagrangian formulation of $\chi$ and $\cd$ electrodynamics}

In this section we show how our system can be described using the language of the effective Lagrangian. While the set of electrodynamics equations derived above allows one to solve any electromagnetic problem for $\cd$ medium, the description in terms of the effective Lagrangian may prove useful for the comparison of our metamaterial system with the systems arising in other contexts, for instance, in condensed matter. Below, we assume that the external sources are absent and consider the reference frame associated with the medium.

We revisit first Lagrangian formulation of axion electrodynamics in the medium, next we consider a case with pure $\cd$ response and finally show how one may construct the Lagrangian description of a metamaterial featuring the combination of $\c$ and $\cd$ responses. Below, we exploit the notations of Sec.~I.

The Lagrangian of axion electrodynamics~\cite{Wilczek87} in the medium reads:
\begin{equation}\label{eq:ChiLagrangian}
    \mathcal{L}_{\chi}=\underbrace{-\frac{1}{16\pi c}\,H^{\mu\nu}F_{\mu\nu}}_{\mathcal{L}_M}-\underbrace{\frac{\chi}{16 \pi c}\,F_{\mu\nu}\tilde F^{\mu\nu}}_{\mathcal{L}_A}\:,
\end{equation}
where the first term describes the usual electrodynamics in the medium, while the second term is the axion contribution. Here, the potential is introduced in the usual way
\begin{equation}\label{eq:Potential}
    F_{\mu\nu}=\partial_\mu A_\nu-\partial_\nu A_\mu\:.
\end{equation}
Such definition of the field tensor immediately yields the second pair of Maxwell's equations. Indeed,
\begin{equation*}
\partial_\mu\tilde{F}^{\mu\nu}=\frac{e^{\mu\nu\alpha\beta}}{2}\,\left(\partial_{\mu}\partial_{\alpha}A_\beta-\partial_\mu\partial_{\beta} A_\alpha\right)=0\,    
\end{equation*}
as Levi-Civita symbol is antisymmetric with respect to the permutation of the indices $\mu$ and $\alpha$ as well as $\mu$ and $\beta$. Hence,
\begin{equation}\label{eq:ChiEq2}
    \partial_\mu \tilde{F}^{\mu\nu}=0\:,
\end{equation}
which is known as Bianchi identity.

The first pair of Maxwell's equations is recovered by writing Euler-Lagrange equations for the potential $A_{\nu}$. For that purpose, we note that
\begin{gather*}
    \frac{\partial F_{\alpha\beta}}{\partial\left(\partial_\mu A_\nu\right)}=\delta^\mu_\alpha\,\delta_\beta^\nu-\delta_\beta^\mu\,\delta_\alpha^\nu\:,\\
    \frac{\partial H^{\alpha\beta}}{\partial\left(\partial_\mu A_\nu\right)}=d^{\mu\alpha}\,d^{\nu\beta}-d^{\nu\alpha}\,d^{\mu\beta}\:.
\end{gather*}
Straightforward calculation then yields:
\begin{equation}\label{eq:chi-tilde-chi-ME1}
\partial_\mu (H^{\mu\nu}+\chi \tilde F^{\mu\nu})=0\:, 
\end{equation}
which is the 4D form of the axion electrodynamics equations [cf. with Eq.~\eqref{eq:Covariant1}].

The problem with the Lagrangian formulation of dual axion electrodynamics is that the second pair of Maxwell's equations is modified, and hence the potential cannot be introduced in the usual way. We overcome this issue by requiring that the first pair of Maxwell's equations plays the role of Bianchi identity. Specifically, we introduce {\it magnetic potential} $\bar{A}_\mu$ defined as
\begin{equation}\label{eq:MagneticPotential}
\tilde{H}_{\mu\nu}=\partial_{\mu}\bar{A}_\nu-\partial_{\nu}\bar{A}_\mu\:,
\end{equation}
which immediately yields
\begin{equation}\label{eq:TchiEq1}
\partial_{\mu} H^{\mu\nu}=0\:.
\end{equation}
This coincides with Eq.~\eqref{eq:Covariant1} in the absence of the sources and axion field. The Lagrangian of $\cd$ electrodynamics is introduced as
\begin{equation}\label{eq:TchiLagrangian}
    \mathcal{L}_{\cd}=\underbrace{-\frac{1}{16\pi c}\,H^{\mu\nu}F_{\mu\nu}}_{\mathcal{L}_M}-\underbrace{\frac{\cd}{16 \pi c}\,H^{\mu\nu}\tilde H_{\mu\nu}}_{\mathcal{L}_D}\:.
\end{equation}
Given the definition of $\tilde{H}_{\mu\nu}$, Eq.~\eqref{eq:MagneticPotential} the rest of the tensors are calculated as
\begin{gather*}
H^{\mu\nu}=-\frac{1}{2}\,e^{\mu\nu\alpha\beta}\,\tilde{H}_{\alpha\beta}\:,\\
F_{\mu\nu}=\left(d^{-1}\right)_{\mu\alpha}\,\left(d^{-1}\right)_{\nu\beta}\,H^{\alpha\beta}\,\\
\tilde{F}^{\mu\nu}=\frac{1}{2}\,e^{\mu\nu\alpha\beta}\,F_{\alpha\beta}\:.
\end{gather*}
Varying the Lagrangian Eq.~\eqref{eq:TchiLagrangian} with respect to the magnetic potentials, we recover
\begin{eqnarray}
    \partial_{\mu}\,\left(\tilde{F}^{\mu\nu}-\cd\,H^{\mu\nu}\right)=0\:,
\end{eqnarray}
which is exactly Eq.~\eqref{eq:Covariant2}.

For the Lagrangian description of the mixed $(\c,\cd)$ case we employ a two-potential approach to electrodynamics \cite{Zwanziger_1971}, which is motivated by the fact that $(\c,\cd)$ case effectively contains both electric and magnetic sources. We write the Lagrangian exactly in the same form as in the conventional medium:
\begin{equation}\label{eq:MixedLagrangian}
\mathcal{L}=-\frac{1}{16\pi c}\,F_{\mu\nu}\,H^{\mu\nu}\:.
\end{equation}
However, the definition of the field tensors is different:
\begin{gather}
F_{\mu\nu}=F^{(1)}_{\mu\nu}-\cd\,F^{(2)}_{\mu\nu}\:,\\
H_{\mu\nu}=\chi\,\tilde{F}^{(1)}_{\mu\nu}+\tilde{F}_{\mu\nu}^{(2)}\:,
\end{gather}
where $F^{(1)}_{\mu\nu}$ and $F^{(2)}_{\mu\nu}$ are defined in terms of the two potentials as
\begin{gather*}
F^{(1,2)}_{\mu\nu}=\partial_\mu A^{(1,2)}_\nu-\partial_\nu A^{(1,2)}_\mu\:.
\end{gather*}

In this formulation, we consider $F_{\mu\nu}$ and $H_{\mu\nu}$ as independent fields and do not impose the connection between them in the form of the constitutive relations. We therefore can vary the Lagrangian independently with respect to the two potentials, which yields the following Euler-Lagrange equations:
\begin{gather}
\partial_\mu\left(H^{\mu\nu}+\chi \tilde{F}^{\mu\nu}\right)=0\:,\\
\partial_{\mu}\left(\tilde{F}^{\mu\nu}-\cd\,H^{\mu\nu}\right)=0\:.
\end{gather}
These equations coincide with Eqs.~\eqref{eq:Covariant1}, \eqref{eq:Covariant2} in the absence of the sources.

\section{VI. Analytical solution for the slab geometry}

In this section we derive the transmission and reflection matrices for a slab possessing effective axion response $\chi$, dual axion response $\cd$ or the combination of both responses simultaneously. The respective permittivity and permeability are $\varepsilon_{1,2}$ and $\mu_{1,2}$.

\begin{figure}[h!]
  \centering
\includegraphics[width=0.6\textwidth]{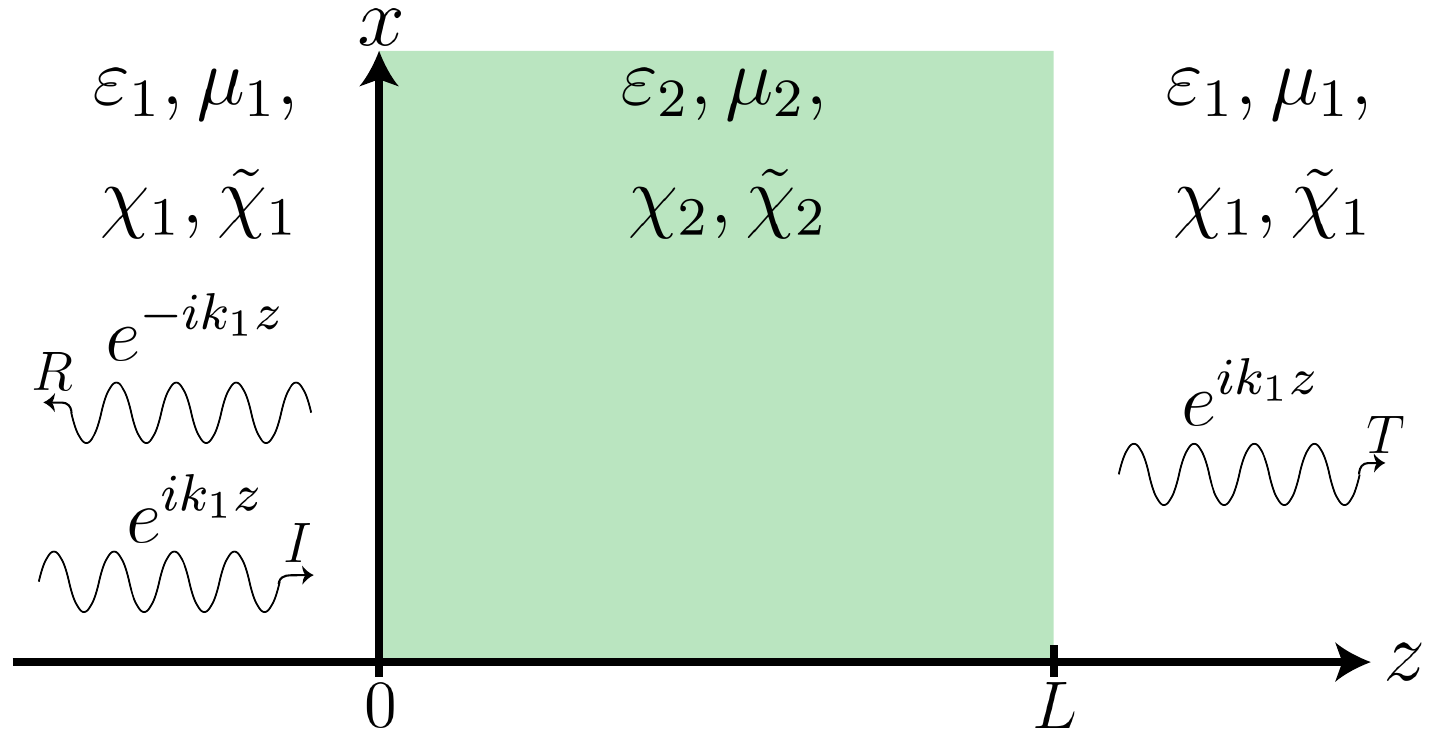}
  \caption{Reflection and transmission of a plane electromagnetic wave from a slab with axion and dual axion response $\chi$ and $\cd$.} 
  \label{fig:mixedslab}
\end{figure}

We assume that the incident plane wave  propagates along the $z$ axis from medium 1 to medium 2 encountering their interface at $z=0$. The transmitted wave propagates through the medium 2 and then hits a second interface, going back to the medium 1. The general case with both $\chi$ and $\cd$ responses is depicted schematically in Fig.~\ref{fig:mixedslab}. Next, we derive the transmission and reflection coefficients for the axion case and the dual case individually. In the last part of this section we compute the reflection and transmission coefficients when both $\chi$ and $\cd$ responses appear simultaneously.

\subsection{Axion case $\cd_1 = \cd_2 =0$}

In the pure axion case the boundary conditions Eq.~\eqref{eq:BC1}-\eqref{eq:BC2} can be presented in the matrix form as follows:
\begin{equation}
    \begin{pmatrix}
        \mathbf{E}_{1t} \\ 
        \bas{z}\times\mathbf{H}_{1t}
    \end{pmatrix} =  \begin{pmatrix}
        \mathbb{1} & 0  \\
        (\chi_2-\chi_1)\,\bas{z}^\times & \mathbb{1}
    \end{pmatrix}
     \begin{pmatrix}
        \mathbf{E}_{2t} \\ 
        \bas{z}\times\mathbf{H}_{2t}
    \end{pmatrix}
\label{timeboundaryaxion012},\end{equation}
where $\bas{z}$ operator is a vector product by $\bas{z}$ expressed as a $2\times 2$ matrix 
\begin{equation}
    \bas{z}^\times=
    \begin{pmatrix}
     0 & -1\\
     1 & 0
    \end{pmatrix}
\end{equation}

This can be further recast as
\begin{equation}
    \begin{pmatrix}
        \vphantom{\sqrt{\dfrac{\varepsilon_1}{\mu_1}}} \mathbf{E}_t \\ \sqrt{\dfrac{\varepsilon_2}{\mu_2}} \mathbf{E}_t
    \end{pmatrix} =  \begin{pmatrix}
        \vphantom{\sqrt{\dfrac{\varepsilon_1}{\mu_1}}}M_{11} & M_{12}  \\
        M_{21} & \vphantom{\sqrt{\dfrac{\varepsilon_1}{\mu_1}}}M_{22}
    \end{pmatrix}
     \begin{pmatrix}
        \vphantom{\sqrt{\dfrac{\varepsilon_1}{\mu_1}}} \mathbf{E}_i  +\mathbf{E}_r \\ \sqrt{\dfrac{\varepsilon_1}{\mu_1}} \Big(\mathbf{E}_i - \mathbf{E}_r \Big)
    \end{pmatrix}
\label{timeboundaryaxion12},\end{equation}
where $M_{11}=\mathbb{1}$, $M_{12}=0$, $M_{22}=\mathbb{1}$, and

\begin{equation}
    M_{21} = (\chi_1-\chi_2) \begin{pmatrix}
        \vphantom{\sqrt{\frac{\varepsilon_1}{\mu_1}}}0 & -1  \\
        1 & \vphantom{\sqrt{\frac{\varepsilon_1}{\mu_1}}}\hphantom{-}0
    \end{pmatrix}.
\end{equation}
The bulk transfer matrix for the propagation inside an axion slab is the same as in isotropic medium and reads
\begin{align}
    \begin{pmatrix}
        \vphantom{\sqrt{\dfrac{\varepsilon_1}{\mu_1}}} \mathbf{E}_t(z) \\ \sqrt{\dfrac{\varepsilon}{\mu}} \mathbf{E}_t(z)
    \end{pmatrix} & = \hat{M}(z) 
    \begin{pmatrix}
        \vphantom{\sqrt{\dfrac{\varepsilon_1}{\mu_1}}} \mathbf{E}_i(0)  +\mathbf{E}_r(0)  \\ \sqrt{\dfrac{\varepsilon'}{\mu'}} \Big(\mathbf{E}_i(0)  - \mathbf{E}_r(0)  \Big)
    \end{pmatrix}, & \hat{M}(z) & = \begin{pmatrix}
        \vphantom{\sqrt{\dfrac{\varepsilon_1}{\mu_1}}} \cos(k' z) & i \sin(k' z) \sqrt{\dfrac{\mu'}{\varepsilon'}} \\
        i \sin(k' z) \sqrt{\dfrac{\varepsilon'}{\mu'}} & \cos(k' z)
    \end{pmatrix}
\label{bulkaxion12}
\end{align}
where $k' = k_0 \sqrt{\mu' \varepsilon'}$ and corresponds to wavevector within the medium with $\varepsilon'$ and $\mu'$. Using the transfer-matrix method we can describe the transmission and reflection of the axion-slab. First, noticing Eq. (\ref{timeboundaryaxion12}), we write the full transfer-matrix

\begin{equation}
    \hat{M} = \hat{M}_{\chi_2\rightarrow \chi_1}\hat{M}(L) \hat{M}_{\chi_2\rightarrow \chi_1} 
\label{axionslabmatrix}\end{equation}
which describes the propagation throughout the axion slab of thickness $L$ with axion field $\chi_2$ from a medium with $\chi_1$.
 
The transmission and reflection matrices can be computed directly from the transfer matrix in Eq. (\ref{axionslabmatrix}) which yields

\begin{align}
    \hat{R} & = -\left[\sqrt{\frac{\varepsilon_1}{\mu_1}} M_{11} -\frac{\varepsilon_1}{\mu_1} M_{12} - M_{21} +\sqrt{\frac{\varepsilon_1}{\mu_1}} M_{22}\right]^{-1} \cdot \left[\sqrt{\frac{\varepsilon_1}{\mu_1}} M_{11} + \frac{\varepsilon_1}{\mu_1} M_{12} - M_{21} -\sqrt{\frac{\varepsilon_1}{\mu_1}} M_{22}\right], \\
    \hat{T} & = \left[ M_{11} + \frac{\varepsilon_1}{\mu_1} M_{12}+ \left( M_{11} - \sqrt{\frac{\varepsilon_1}{\mu_1}} M_{12}\right)\right] \cdot \hat{R},
\end{align}
The components of the reflection and transmission matrices for the axion slab are

\begin{align}
    t_{11} \hphantom{-} = \hphantom{-}t_{22} & = \frac{2 i}{g(\chi)} \sqrt{\frac{\varepsilon_1\varepsilon_2}{\mu_1 \mu_2}}, \label{Txx:axionslab}\\
    t_{12} \hphantom{-} = \hphantom{-}t_{21} & = 0, \label{Txy:axionslab}\\
    r_{11} \hphantom{-} = \hphantom{-}r_{22} & =  \frac{1}{g(\chi)} \left( \frac{\varepsilon_1}{\mu_1} - \frac{\varepsilon_2}{\mu_2} - \chi^2\right) \sin(k_2 L), \label{Rxx:axionslab} \\
    r_{12} \hphantom{-} = - r_{21} & = - \frac{2}{g(\chi)}\sqrt{\dfrac{\varepsilon_1 }{\mu_1}} \chi \sin(k_2 L),\label{Rxy:axionslab}
\end{align}
where 
\begin{equation}
    g(\chi) = 2i\sqrt{\frac{\varepsilon_1\varepsilon_2}{\mu_1 \mu_2}} \cos(k_2 L) +\left( \frac{\varepsilon_1}{\mu_1} + \frac{\varepsilon_2}{\mu_2} + \chi^2\right) \sin(k_2 L), 
\end{equation}
$\chi= \chi_1 - \chi_2$, and $k_2 = k_0 \sqrt{\mu_2 \varepsilon_2}$.

\subsection{Dual axion case $\c_1 = \c_2 =0$}

The dual axion case uses the same equations with the only difference in the boundary conditions which can be presented in the form 
\begin{equation}
    \begin{pmatrix}
        \vphantom{\sqrt{\dfrac{\varepsilon_1}{\mu_1}}} \mathbf{E}_t \\ \sqrt{\dfrac{\varepsilon_2}{\mu_2}} \mathbf{E}_t
    \end{pmatrix} =  \begin{pmatrix}
        \vphantom{\sqrt{\dfrac{\varepsilon_1}{\mu_1}}}M_{11} & M_{12}  \\
        M_{21} & \vphantom{\sqrt{\dfrac{\varepsilon_1}{\mu_1}}}M_{22}
    \end{pmatrix}
     \begin{pmatrix}
        \vphantom{\sqrt{\dfrac{\varepsilon_1}{\mu_1}}} \mathbf{E}_i  +\mathbf{E}_r \\ \sqrt{\dfrac{\varepsilon_1}{\mu_1}} \Big(\mathbf{E}_i - \mathbf{E}_r \Big)
    \end{pmatrix}
\label{timeboundaryaxion142},\end{equation}
where $M_{11}=\mathbb{1}$, $M_{21}=0$, $M_{22}=\mathbb{1}$, and

\begin{equation}
    M_{12} = (\cd_1-\cd_2) \begin{pmatrix}
        \vphantom{\sqrt{\frac{\varepsilon_1}{\mu_1}}}0 & -1  \\
        1 & \vphantom{\sqrt{\frac{\varepsilon_1}{\mu_1}}}\hphantom{-}0
    \end{pmatrix}.
\end{equation}
The bulk matrix propagation remains the same. Now, the full transfer matrix describing the propagation throughout a dual axion slab is given by 

\begin{equation}
    \hat{M} = \hat{M}_{\cd_2\rightarrow \cd_1}\hat{M}(L) \hat{M}_{\cd_2\rightarrow \cd_1} 
\label{dualaxionslabmatrix}\end{equation}
Using the same formulation to derive the reflection and transmission matrix components for the dual axion slab we obtain

\begin{align}
    t_{11} \hphantom{-} = \hphantom{-}t_{22} & = \frac{2 i}{\tilde{g}(\cd)} \sqrt{\frac{\mu_1 \mu_2}{\varepsilon_1\varepsilon_2}}, \label{Txx:dualaxionslab}\\
    t_{12} \hphantom{-} = \hphantom{-}t_{21} & = 0, \label{Txy:dualaxionslab}\\
    r_{11} \hphantom{-} = \hphantom{-}r_{22} & =  \frac{1}{\tilde{g}(\cd)} \left(  \frac{\mu_2}{\varepsilon_2} - \frac{\mu_1}{\varepsilon_1} + \cd^2\right) \sin(k_2 L), \label{Rxx:dualaxionslab} \\
    r_{12} \hphantom{-} = - r_{21} & = \frac{2}{\tilde{g}(\cd)}\sqrt{\dfrac{\mu_1 }{\varepsilon_1}} \cd \sin(k_2 L),\label{Rxy:dualaxionslab}
\end{align}
where 
\begin{equation}
    \tilde{g}(\cd) = 2i\sqrt{\frac{\mu_1 \mu_2}{\varepsilon_1\varepsilon_2}} \cos(k_2 L) +\left( \frac{\mu_1}{\varepsilon_1} + \frac{\mu_2}{\varepsilon_2} + \cd^2\right) \sin(k_2 L), 
\end{equation}
and $\cd= \cd_1 - \cd_2$.

It should be stressed that both solutions predict Kerr rotation for the reflected wave and no Faraday rotation for the transmitted light. Moreover, it is straightforward to verify that Eqs.~\eqref{Txx:dualaxionslab}-\eqref{Rxy:dualaxionslab} are converted to Eqs.~\eqref{Txx:axionslab}-\eqref{Rxy:axionslab} using the mapping Eqs.~\eqref{eq:eps_eff}-\eqref{eq:chi_eff} and setting $\chi_1=\cd_1=0$. This illustrates that the scattering experiments typically used to diagnose the effective axion fields are insufficient to distinguish $\chi$ and $\cd$ responses.

\subsection{Mixed axion case}
In the presence of both axion and dual axion responses the boundary transfer matrix takes the form
\begin{equation}
    \begin{pmatrix}
        \vphantom{\sqrt{\dfrac{\varepsilon_1}{\mu_1}}} \mathbf{E}_t \\ \sqrt{\dfrac{\varepsilon_2}{\mu_2}} \mathbf{E}_t
    \end{pmatrix} = \frac{1}{1+\chi \cd} \begin{pmatrix}
        \vphantom{\sqrt{\dfrac{\varepsilon_1}{\mu_1}}} \mathbb{1} & \cd e^\times_z  \\
        \chi e^\times_z & \vphantom{\sqrt{\dfrac{\varepsilon_1}{\mu_1}}} \mathbb{1}
    \end{pmatrix}
     \begin{pmatrix}
        \vphantom{\sqrt{\dfrac{\varepsilon_1}{\mu_1}}} \mathbf{E}_i  +\mathbf{E}_r \\ \sqrt{\dfrac{\varepsilon_1}{\mu_1}} \Big(\mathbf{E}_i - \mathbf{E}_r \Big)
    \end{pmatrix}
\label{timeboundaryaxion142},\end{equation}
where 

\begin{equation}
  e^\times_z =  \begin{pmatrix}
        \vphantom{\sqrt{\frac{\varepsilon_1}{\mu_1}}}0 & -1  \\
        1 & \vphantom{\sqrt{\frac{\varepsilon_1}{\mu_1}}}\hphantom{-}0
    \end{pmatrix}.
\end{equation}
As done before, we compute the full transfer matrix by the multiplication of the boundary transfer matrices and the bulk transfer matrix. The transmission and reflection matrix components for the mixed axion and dual axion case read
\begin{align}
    t_{11} \hphantom{-} = \hphantom{-}t_{22} & = \frac{2 i}{\Delta(\chi,\cd)} \Big(1+ \chi \tilde{\chi}\Big)\sqrt{\frac{\mu_1 \mu_2}{\varepsilon_1\varepsilon_2}}, \label{Txx:mixedaxionslab}\\
    t_{12} \hphantom{-} = \hphantom{-}t_{21} & = 0, \label{Txy:mixedaxionslab}\\
    r_{11} \hphantom{-} = \hphantom{-}r_{22} & =  \frac{1}{\Delta(\chi,\cd)} \left(  \frac{\mu_2}{\varepsilon_2} - \frac{\mu_1}{\varepsilon_1} -\frac{\mu_1 \mu_2}{\varepsilon_1\varepsilon_2} \chi^2+ \cd^2\right) \sin(k_2 L), \label{Rxx:mixedaxionslab} \\
    r_{12} \hphantom{-} = - r_{21} & = \frac{2}{\Delta(\chi,\cd)}\sqrt{\dfrac{\mu_1 }{\varepsilon_1}} \bigg(\cd  -\frac{\mu_2}{\eps_2}\chi\bigg)\sin(k_2 L),\label{Rxy:mixedaxionslab}
\end{align}
where 
\begin{equation}
    \Delta(\chi,\cd) = 2i\sqrt{\frac{\mu_1 \mu_2}{\varepsilon_1\varepsilon_2}} \bigg(1 + \chi \cd \bigg)\cos(k_2 L) +\left( \frac{\mu_1}{\varepsilon_1} + \frac{\mu_2}{\varepsilon_2} + \frac{\mu_1 \mu_2}{\varepsilon_1\varepsilon_2}\chi^2+ \cd^2\right) \sin(k_2 L), 
\end{equation}

As can be readily verified, Eqs.~\eqref{Txx:mixedaxionslab}-\eqref{Rxy:mixedaxionslab} are converted to Eqs.~\eqref{Txx:axionslab}-\eqref{Rxy:axionslab} using a more general form of mapping Eqs.~\eqref{eq:MappingG1}-\eqref{eq:MappingG3}. As expected, transmission and reflection data does not allow to discriminate the contributions stemming from $\chi$ and $\cd$ responses.

\section{VII. Transfer matrix description of a multilayered metamaterial}

In this section we present the transfer matrix method applied to the multilayered metamaterial structure and validate the description of the metamaterial in terms of effective axion and dual axion fields.

As a first step, we consider a one-dimensional (1D) photonic crystal (PhC) composed of two periodic gyrotropic layers, A and B possessing gyrotropic permittivity and permeability tensors of the following form
\begin{equation}
\hat{\eps}(z + D) =\hat{\eps}(z) =\begin{pmatrix}
\eps & i\,g(z) & 0\\
-i\,g(z) & \eps & 0\\
0 & 0 & \eps
\end{pmatrix},\quad
\hat{\mu}(z + D) = \hat{\mu}(z) =\begin{pmatrix}
\mu & i\,f(z) & 0\\
-i\,f(z) & \mu & 0\\
0 & 0 & \mu
\end{pmatrix}.
\end{equation}
where $D=d_A+d_B$ is the metamaterial period, and $d_\alpha$ is the thickness of the $\alpha$ layer. 

While the verification of the effective medium picture for the layered axion metamaterial by the transfer-matrix method has been done in Ref.~\cite{Shaposhnikov2023}, here we focus on the two complementary cases: the dual-axion metamaterial and metamaterial possessing the combination of axion and dual axion responses.

\subsection{Dual axion metamaterial}
In order to generate only the dual axion response, it is sufficient to introduce only magnetic gyrotropy in the layers. The constitutive relations for a single layer $\alpha = A, B$ read
\begin{align}
    B^i_\alpha & = \mu_\alpha^{ij} \; H_\alpha^j, &  \mu_\alpha^{ij} & =  
    \begin{pmatrix}
        \eps_\alpha & i f_\alpha &  0\\
        -if_\alpha & \eps_\alpha &0  \\
         0 & 0  &\mu_\alpha  \\
    \end{pmatrix}
\end{align}
where we assume $f_A = -f_B = f$. Using a chiral basis, $\mathbf{H}_\alpha = H_\alpha \,\hat{\vc{x}}+ i\eta H_\alpha \,\hat{\vc{y}}$, where $\eta = \pm 1$ for right- and left-handed polarization, we can write the $\vc{B}$ fields as

\begin{align}
    B_{\alpha,\eta} & = \mu_{\alpha,\eta} H_{\alpha,\eta}, & \mu_{\alpha,\eta} & =  \mu_\alpha-\eta f_\alpha.
\end{align}
Therefore, we can construct the electric field propagating in a layer $\alpha$ as~\cite{Araujo2021Nov}

\begin{equation}
    \mathbf{E}_{\alpha,\eta} = E_{\alpha,\eta}\; e^{i k_{\alpha,\eta} t} \hat{{v}}_\eta + E'_{\alpha,\eta} \; e^{-i k_{\alpha,\eta} t} \hat{{v}}_\eta
\label{Echiralbasis}\end{equation}
and
\begin{equation}
    \mathbf{H}_{\alpha,\eta} = -i\eta \sqrt{\dfrac{\varepsilon_{\alpha}}{\mu_{\alpha,\eta}}} E_{\alpha,\eta}\; e^{i k_{\alpha,\eta} t} \hat{{v}}_\eta + i\eta \sqrt{\dfrac{\varepsilon_{\alpha}}{\mu_{\alpha,\eta}}} E'_{\alpha,\eta}\; e^{-i k_{\alpha,\eta} t} \hat{{v}}_\eta,
\label{Hchiralbasis}\end{equation}
where 
\begin{equation}
    k_{\alpha,\eta} = k_0 \sqrt{\varepsilon_{\alpha} \mu_{\alpha,\eta}},
\label{dispersionrelationgyrolayer}\end{equation}
is the wave number, $k_0 = \omega/c$ is a vacuum wave number, $E_{\alpha,\eta}$ and $E'_{\alpha,\eta}$ are the amplitudes of the wave travelling to the right (left). Here, $\hat{v}_\eta= \hat{v}_\pm$ vector is given by
\begin{equation}
    \hat{v}_\eta = \frac{1}{\sqrt{2}}
    \begin{pmatrix}
        1 \\
        i \eta \\
        0 
    \end{pmatrix}.
\end{equation}
By applying the conventional boundary conditions, i.e., the continuity of the tangential components of the $\vc{H}$ and $\vc{E}$ fields at the boundaries of the layers, to Eqs. (\ref{Echiralbasis}),(\ref{Hchiralbasis}) we obtain
\begin{align}
    E_A + E'_A & = E_B + E'_B,\label{boundarygyrotropictime1} \\
    \xi_A E_A -  \xi_A E'_A & =  \xi_B  E_B - \xi_B  E'_B, \label{boundarygyrotropictime2}
\end{align}
where $\xi_\alpha = - i\eta \sqrt{\varepsilon_{\alpha} / \mu_{\alpha,\eta}}$. From the latter set of equations, (\ref{boundarygyrotropictime1}),(\ref{boundarygyrotropictime2}) we can write the $2 \times 2$ interface matrix connecting the fields in layers $A$ to $B$:
\begin{align}
    \begin{pmatrix}
        E_B\\
        E'_B
    \end{pmatrix} & = M_{AB} 
    \begin{pmatrix}
        E_A\\
        E'_A
    \end{pmatrix}, & M_{AB} & = \frac{1}{2}
    \begin{pmatrix}
        1 + \dfrac{Y_A}{Y_B} &  1 - \dfrac{Y_A}{Y_B} \\
        1 - \dfrac{Y_A}{Y_B} &  1 + \dfrac{Y_A}{Y_B}
    \end{pmatrix}.
\end{align}
We adopt a basis with the amplitude vector constructed as $(E_{\alpha,+},E'_{\alpha,+},E_{\alpha,-},E'_{\alpha,-})^{T}$, where $+(-)$ describes the right- (left-) handed polarization. The $4\times 4$ interface matrix is given by

\begin{equation}
    \begin{pmatrix}
         \vphantom{\dfrac{Y_{A,+}}{Y_{B,+}}} E_{B,+} \\
         \vphantom{\dfrac{Y_{A,+}}{Y_{B,+}}} E'_{B,+} \\
         \vphantom{\dfrac{Y_{A,+}}{Y_{B,+}}} E_{B,-} \\
         \vphantom{\dfrac{Y_{A,+}}{Y_{B,+}}} E'_{B,-}
    \end{pmatrix} =
    \frac{1}{2} \begin{pmatrix}
         1 + \dfrac{Y_{A,+}}{Y_{B,+}} &  1 - \dfrac{Y_{A,+}}{Y_{B,+}} & 0 & 0 \\
         1 - \dfrac{Y_{A,+}}{Y_{B,+}} & 1 + \dfrac{Y_{A,+}}{Y_{B,+}}&  0 & 0 \\
         0 & 0 & 1 + \dfrac{Y_{A,-}}{Y_{B,-}} &  1 - \dfrac{Y_{A,-}}{Y_{B,-}} \\
         0 & 0 & 1 - \dfrac{Y_{A,-}}{Y_{B,-}} &  1 + \dfrac{Y_{A,-}}{Y_{B,-}}
    \end{pmatrix}
    \begin{pmatrix}
         \vphantom{\dfrac{Y_{A,+}}{Y_{B,+}}} E_{A,+} \\
         \vphantom{\dfrac{Y_{A,+}}{Y_{B,+}}} E'_{A,+} \\
         \vphantom{\dfrac{Y_{A,+}}{Y_{B,+}}} E_{A,-} \\
         \vphantom{\dfrac{Y_{A,+}}{Y_{B,+}}} E'_{A,-}
    \end{pmatrix},
\end{equation}
where we define $Y_{\alpha, \eta} = \sqrt{\varepsilon_{\alpha} / \mu_{\alpha,\eta}}$. Finally, for the case of an electromagnetic wave propagating inside the medium $\alpha=A,B$, with thickness $d_\alpha$ and wavevector $k_{\alpha,\eta}$, the $4\times 4$ bulk-propagation matrix is given by

\begin{equation}
    M_\alpha = M_\alpha(d_\alpha) = 
    \begin{pmatrix}
        e^{i k_{\alpha,+} d_\alpha} & 0 & 0 & 0 \\
        0 & e^{-i k_{\alpha,+} d_\alpha} & 0 & 0 \\
        0 & 0 & e^{i k_{\alpha,-} d_\alpha} & 0 \\
        0 & 0 & 0 & e^{-i k_{\alpha,-} d_\alpha}
    \end{pmatrix}
\end{equation}
The resultant transfer matrix describing the gyrotropic multilayered  structure with even number of $2N$ layers then reads

\begin{equation}
    M_{\text{slab}} = M_{B-\text{air}} M_{AB} (M_{BA} M_{B} M_{AB} M_{A})^{2N} M_{\text{air}-A}  = 
    \begin{pmatrix}
        M^+_{\text{slab}} & 0 \\
        0 & M^-_{\text{slab}}
    \end{pmatrix}
\end{equation}
This matrix connects incident $E^{in}$, reflected $E^r$ and transmitted $E^t$ fields in the following way:
\begin{align}
    \begin{pmatrix}
        E^t_\eta \\
        0
    \end{pmatrix} & = M^\eta_{\text{slab}} 
    \begin{pmatrix}
        E^i_\eta \\
        E^r_\eta
    \end{pmatrix}, & M^\eta_{\text{slab}} & =
    \begin{pmatrix}
        M^\eta_{11} & M^\eta_{12} \\
        M^\eta_{21} & M^\eta_{22}
    \end{pmatrix}
\end{align}
Using this, we deduce the amplitudes of the transmitted and reflected wave for both polarizations 
\begin{align}
    E^t_\eta & =\frac{\text{det}(M^\eta)}{M^\eta_{22}} E^i_\eta &  E^r_\eta & = -\frac{M^\eta_{21}}{M^\eta_{22}} E^i_\eta.
\end{align}
If the incident wave is polarized along the $x$ axis, the amplitudes of reflected and transmitted waves read
\begin{eqnarray}
    \mathbf{E}^t & = & \frac{1}{2}\left(\frac{\text{det}(M^+)}{M^+_{22}} + \frac{\text{det}(M^-)}{M^-_{22}} \right) E^i \, \hat{\vc{x}} + \frac{i}{2}\left(\frac{\text{det}(M^+)}{M^+_{22}} - \frac{\text{det}(M^-)}{M^-_{22}} \right) E^i \, \hat{\vc{y}} \\
    \mathbf{E}^r & = & - \frac{1}{2}\left( \frac{M^+_{21}}{M^+_{22}} + \frac{M^-_{21}}{M^-_{22}} \right) E^i \, \hat{\vc{x}} - \frac{i}{2}\left( \frac{M^+_{21}}{M^+_{22}} - \frac{M^-_{21}}{M^-_{22}} \right) E^i \, \hat{\vc{y}}
\end{eqnarray}
which yields the co- and cross-polarized transmission and reflections in terms of the components of the transfer-matrix for the full multilayered slab

\begin{figure}[hb]
\centering
\includegraphics[width=0.98 \textwidth]{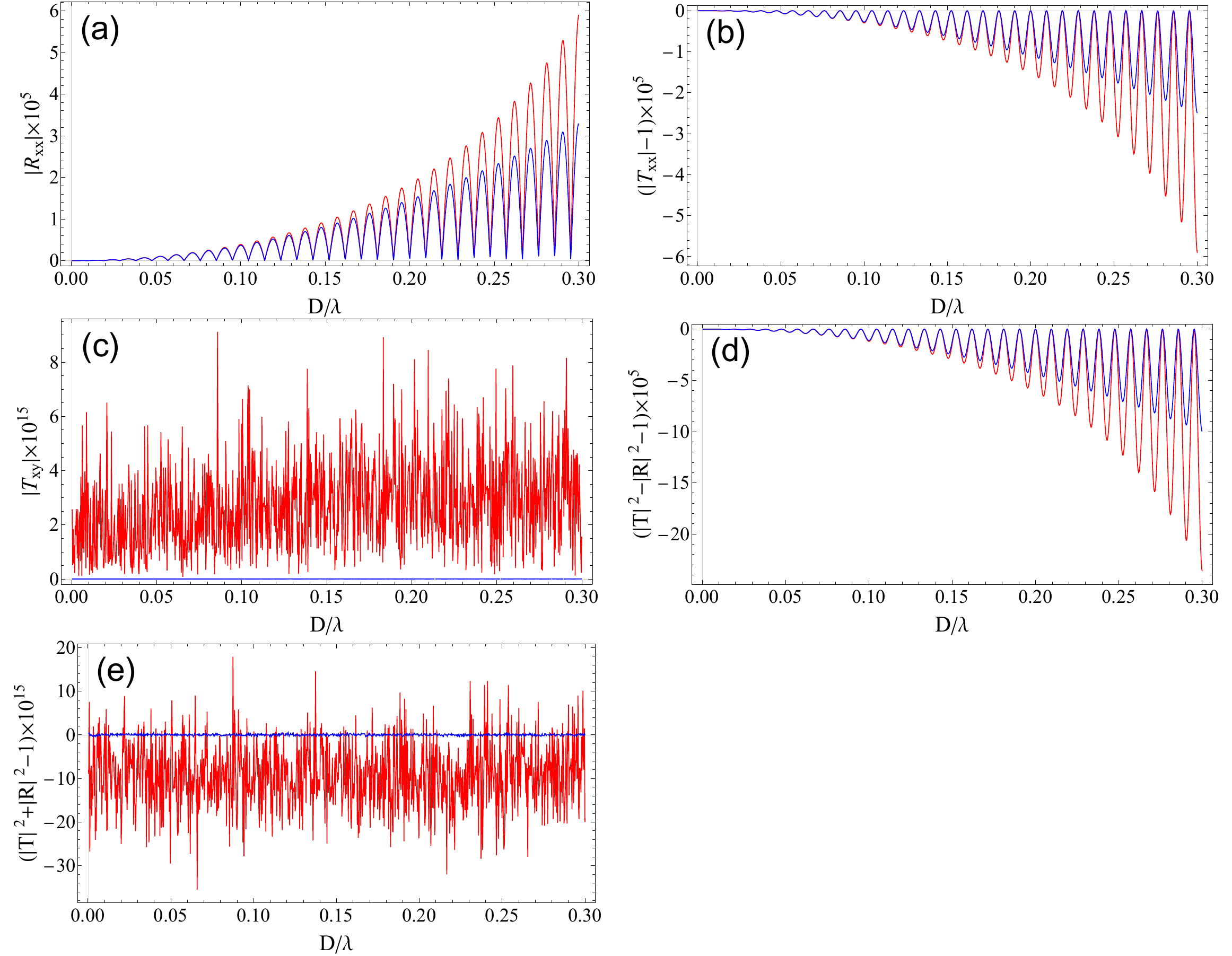}
\caption{Validation of the effective dual axion response. (a-e) Blue lines correspond to the effective medium approximation and red lines to the results of the transfer-matrix method applied to the multilayered structure. Panel (a) corresponds to the co-polarized reflection, (b) shows the co-polarized transmission coefficient, panel (c) illustrates the cross-polarized transmission which is essentially zero, panel (d) corroborates that the momentum $|T|^2-|R|^2 \neq 1$ is not conserved, and panel (e) shows that conservation of energy $|T|^2+|R|^2 = 1$.  
Parameters: number of periods $N = 50$, $\varepsilon_1 = 1, \mu_1=1, \varepsilon_2 = 1.05, \mu_2=1.05, f=0.015$.}
\label{tchi_verification1}
\end{figure}


\begin{align}
    T_{xx} & = \frac{1}{2}\left(\frac{\text{det}(M^+)}{M^+_{22}} + \frac{\text{det}(M^-)}{M^-_{22}} \right), & T_{xy} & = \frac{i}{2}\left(\frac{\text{det}(M^+)}{M^+_{22}} - \frac{\text{det}(M^-)}{M^-_{22}} \right), \label{T:photoniccrystal}\\
    R_{xx} & =- \frac{1}{2}\left( \frac{M^+_{21}}{M^+_{22}} + \frac{M^-_{21}}{M^-_{22}} \right), & R_{xy} & = - \frac{i}{2}\left( \frac{M^+_{21}}{M^+_{22}} -\frac{M^-_{21}}{M^-_{22}} \right).\label{R:photoniccrystal}
\end{align}
In order to perform numerical validation of the effective medium description we consider a photonic crystal (PhC) composed of 50 unit cells. Each unit cell consists of two gyrotropic layers of the same thickness $D/2$ and the same magnitude $f_0= 0.015$ but opposite directions of magnetization. The permittivity and permeability of the gyrotropic layers are $\mu_2 = \varepsilon_2 = 1.05$, and we assume that the PhC is placed in the vacuum $\mu_1 = \varepsilon_1 = 1$.

The effective axion response describes the boundary conditions, and the boundary conditions depend on the termination of the layers~\cite{Gorlach2020}. If the first layer has a positive gyrotropy, $f_0$, then the effective dual axion response, $\cd_\text{eff}$ and the effective permeability $\mu_\text{eff} $ are given by
\begin{align}
    \tilde{\chi}_\text{eff} & = \frac{\pi}{2} \, f_0\,\xi, & \mu_\text{eff} & =  \mu_2 + \frac{\pi^2}{12} f_0^2  \, \xi^2,
\label{effectivedualaxionf0}
\end{align}
with $\varepsilon_\text{eff} = \varepsilon_2$. Here, $\xi(\lambda) = k_z/Q = D /\lambda$ is the period-to-wavelength ratio, and $\lambda$ is the vacuum wavelength. The transmission and reflection coefficients for the dual axion slab can be computed analytically as given by Eqs.~(\ref{Txx:dualaxionslab})-(\ref{Rxy:dualaxionslab}). The optical path inside the slab is defined as:
\begin{equation}
k_2 L = \frac{2\pi}{\sqrt{\mu_{\text{eff}} \, \varepsilon_{\text{eff}}}} N \xi,
\end{equation}
where $N = 50$ represents the number of periods in the multilayered structure. We plot the transmission and reflection coefficients by varying the period-to-wavelength ratio $\xi$ from $0$ to $0.3$. The results for the numerical validation are shown in Fig.~\ref{tchi_verification1} and show adequate correspondence for the effective medium approximation $D/\lambda \lesssim 0.15$ for most of the components. The key parameters to track are cross-polarized transmission $T_{xy}$ and cross-polarized reflection $R_{xy}$. The former vanishes in all three cases (axion, dual axion and combination of both responses), while the cross-polarized reflection allows to determine the magnitude of those responses.

\subsection{Axion and dual axion metamaterial}
In the case when there is both electric and magnetic gyrotropy in the layers, the system features both axion and dual axion effective responses. The constitutive relations for a single layer $\alpha = A, B$ are
\begin{align}
    B^i_\alpha & = \mu_\alpha^{ij} \; H_\alpha^j, &  \mu_\alpha^{ij} & =  
    \begin{pmatrix}
        \eps_\alpha & i f_\alpha &  0\\
        -if_\alpha & \eps_\alpha &0  \\
         0 & 0  &\mu_\alpha  \\
    \end{pmatrix},
\end{align}
\begin{align}
    D^i_\alpha & = \eps_\alpha^{ij} \; E_\alpha^j, &  \eps_\alpha^{ij} & =  
    \begin{pmatrix}
        \eps_\alpha & i g_\alpha &  0\\
        -ig_\alpha & \eps_\alpha &0  \\
         0 & 0  &\mu_\alpha  \\
    \end{pmatrix},
\end{align}
where we will consider $g_A = g = -g_B$. Using a chiral basis, $\mathbf{E}_\alpha = E_\alpha \,\hat{\vc{x}}+ i\eta E_\alpha \,\hat{\vc{y}}$, where $\eta = \pm 1$ for right- and left-handed polarization, we can write the $D$ fields as
\begin{align}
    D_{\alpha,\eta} & = \eps_{\alpha,\eta} E_{\alpha,\eta}, & \eps_{\alpha,\eta} & =  \varepsilon_\alpha-\eta g_\alpha, \\
    B_{\alpha,\eta} & = \mu_{\alpha,\eta} H_{\alpha,\eta}, & \mu_{\alpha,\eta} & =  \mu_\alpha-\eta f_\alpha
\end{align}
Therefore, just as we did previously, we can construct the electric field propagating in a layer $\alpha$ as
\begin{equation}
    \mathbf{E}_{\alpha,\eta} = E_{\alpha,\eta}\; e^{i k_{\alpha,\eta} t} \hat{{v}}_\eta + E'_{\alpha,\eta} \; e^{-i k_{\alpha,\eta} t} \hat{{v}}_\eta
\label{Echiralbasisdual}\end{equation}
and
\begin{equation}
    \mathbf{H}_{\alpha,\eta} = -i\eta \sqrt{\dfrac{\varepsilon_{\alpha,\eta}}{\mu_{\alpha,\eta}}} E_{\alpha,\eta}\; e^{i k_{\alpha,\eta} t} \hat{{v}}_\eta + i\eta \sqrt{\dfrac{\varepsilon_{\alpha,\eta}}{\mu_{\alpha,\eta}}} E'_{\alpha,\eta}\; e^{-i k_{\alpha,\eta} t} \hat{{v}}_\eta,
\label{Hchiralbasisdual}\end{equation}
where 

\begin{equation}
    k_{\alpha,\eta} = k_0 \sqrt{\varepsilon_{\alpha,\eta} \mu_{\alpha,\eta}},
\label{dispersionrelationgyrolayerdual}\end{equation}
is the wavevector and $k_0 = \omega/c$ is a vacuum wave number. The computational procedure is the same as in the dual axion case; the only adjustment is the replacement:
\begin{equation}
    \eps_\alpha \rightarrow \eps_{\alpha,\eta} = \eps_\alpha - \eta g_\alpha.
\end{equation}
From now on we can use all the previously developed machinery.

\begin{figure}[h!]
\centering
\includegraphics[width=0.98 \textwidth]{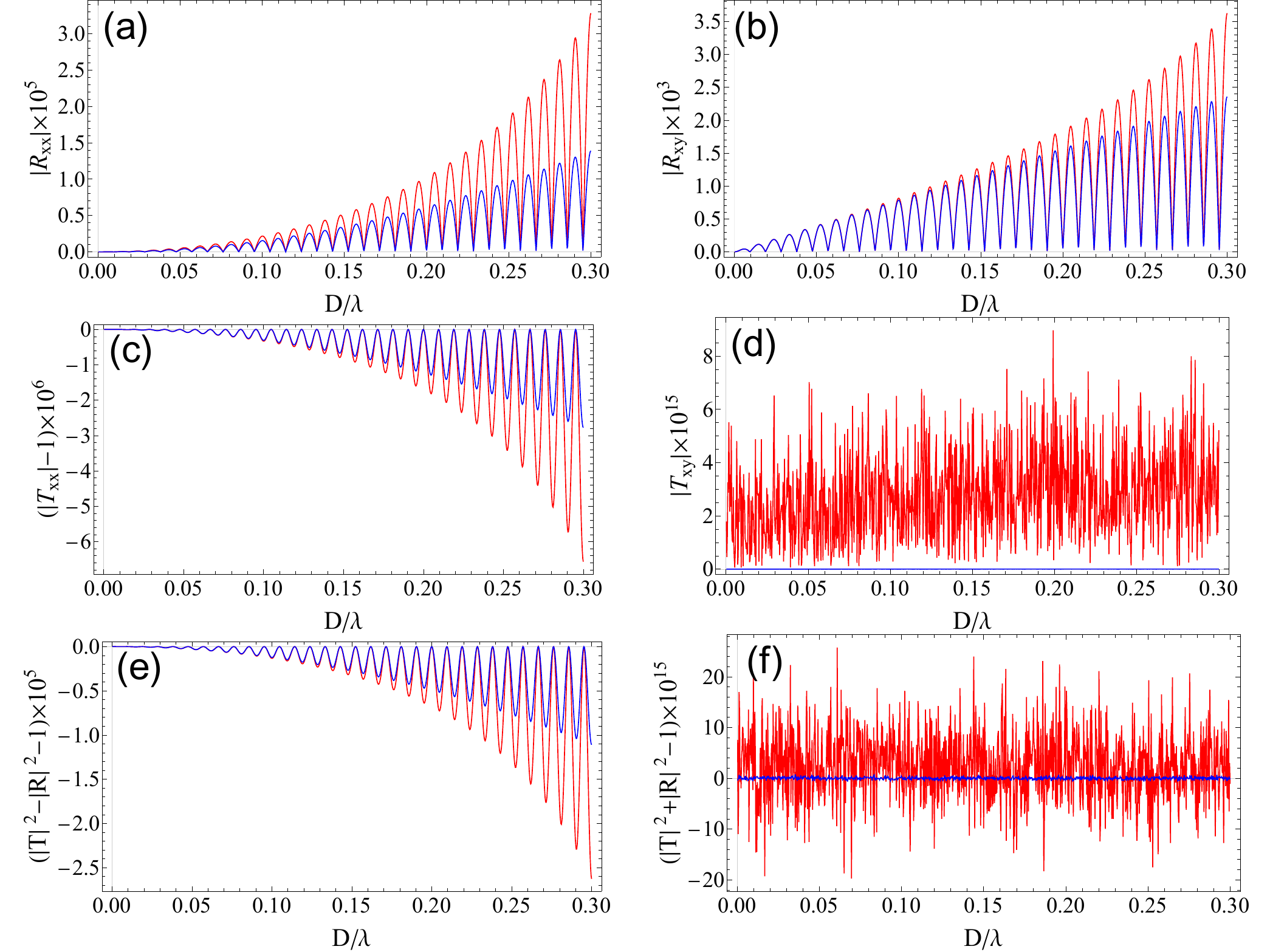}
\caption{Validation of the effective axion plus the effective dual axion response. (a-f) Blue lines correspond to the effective medium approximation and red lines to the results of the transfer-matrix method applied to the multilayered structure. Panel (a) corresponds to the co-polarized reflection, panel (b) shows the cross-polarized reflection coefficient, panel (c) illustrates the co-polarized transmission, panel (d) illustrates the cross-polarized transmission which is essentially zero, panel (e) corroborates that the momentum $|T|^2-|R|^2 \neq 1$ is not conserved, and panel (e) shows that energy $|T|^2+|R|^2 = 1$ is conserved.  
Parameters: number of periods $N = 50$, $\varepsilon_1 = 1, \mu_1=1, \varepsilon_2 = 1.05, \mu_2=1.05, g=0.01, f=0.015$}
\label{chi+tchi_verification}
\end{figure}

The results of numerical validation are depicted in Fig.~\ref{chi+tchi_verification} and also show that the effective medium approximation remains adequate provided $D/\lambda \lesssim 0.15$. The agreement is good both for co- and cross-polarized reflection and transmission coefficients.

\section{VIII. Distinguishing $\chi$ and $\cd$: static dipole in a sphere}

\begin{figure}[t!]
\centering
    \includegraphics[scale=0.25]{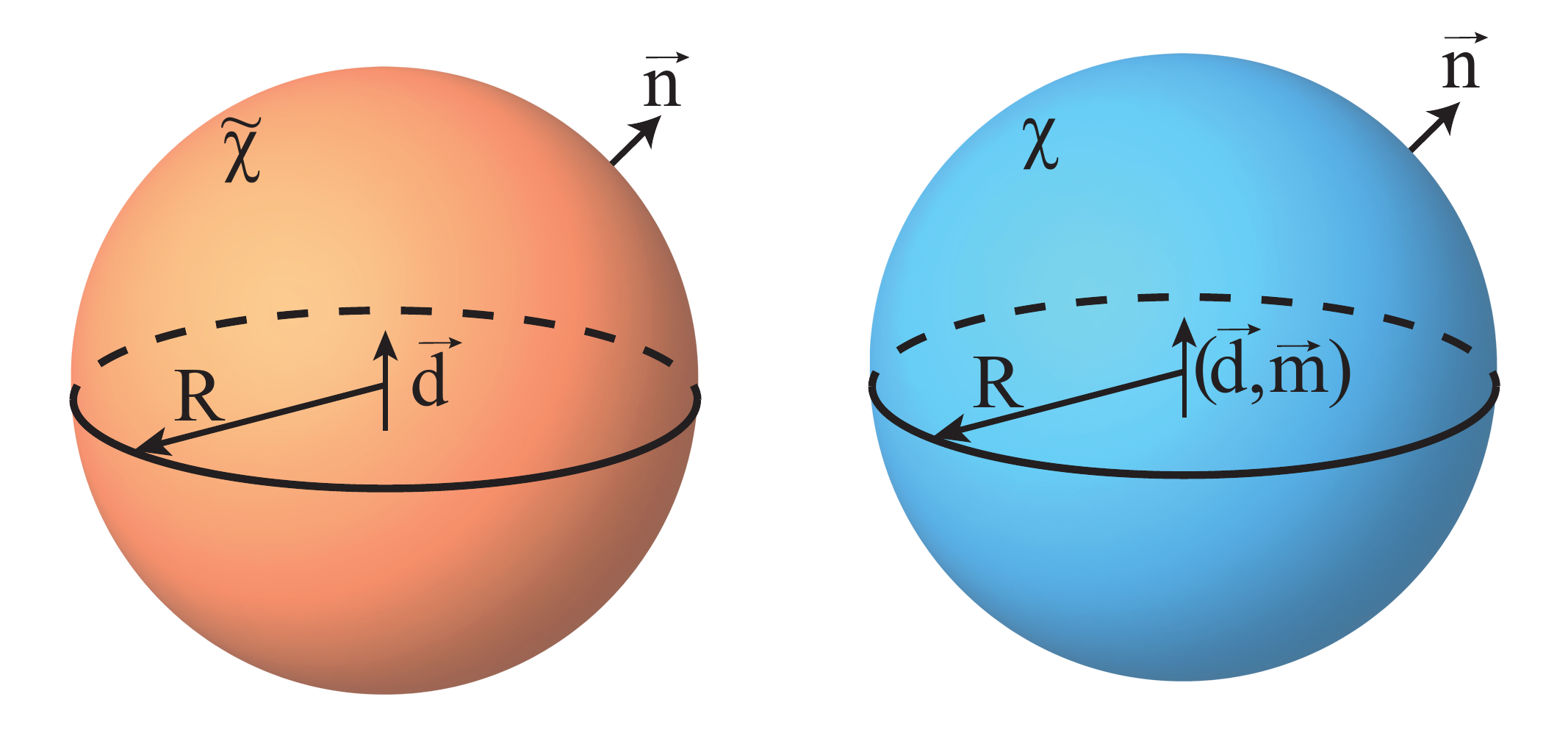}
\caption{Geometry of the problem: a sphere with $\c$ or $\cd$ response is excited by the static dipole sources.}
\label{fig:chi_tchi_d_dm}
\end{figure}

To provide a clear-cut distinction between $\chi$ and $\cd$ responses we consider the following geometry. A sphere of radius $R$ possesses permittivity $\eps$, permeability $\mu$ and axion response $\chi$ or dual axion response $\cd$ being placed in vacuum. We study the excitation of this sphere by the probe static electric dipole $\vc{d}_0$, static magnetic dipole $\vc{m}_0$ or their combination placed in the center of the sphere (Fig.\ref{fig:chi_tchi_d_dm}). We calculate the fields outside of the sphere extracting from those fields effective electric and magnetic dipole moments of the system. Solving this problem analytically, we illustrate that $\chi$ and $\cd$ responses can in principle be distinguished {\it in an experimental situation.}


We start with a dual axion sphere and an electric dipole put inside. We present the fields inside as a superposition of the electric dipole field produced by the probe dipole $\vc d_0$ and some constant boundary-induced field, containing both electric and magnetic components:
\begin{gather}
    \vc{E}_{\text{in}}=\frac{1}{\eps r^3}[3(\vc {n}\cdot\vc {d}_0)\vc {n}-\vc {d}_0]+\vc {E}_0\:, \quad
    \vc{B}_{\text{in}}=\vc B_0\:.
\end{gather}
The fields outside are a superposition of electric and magnetic dipoles, as it was calculated in Ref.~\cite{Seidov2023} for an axion case.
\begin{gather}
    \vc{E}_{\text{out}}=\frac{1}{r^3}[3(\vc{n}\cdot\vc{d}_{\cd})\vc{n}-\vc{d}_{\cd}] \:, \\
    \vc{B}_{\text{out}}=\frac{1}{r^3} [3(\vc{n}\cdot\vc{m}_{\cd})\vc{n}-\vc{m}_{\cd}] \:.
\end{gather}
Note that for the chosen geometry unit radius-vector $\vc n\equiv \vc r /r$ coincides with the surface normal of the medium and therefore we denote them by the same symbol $\vc n$. We use boundary conditions for the dual axion medium 
    \begin{gather}
        [\vc{B}]_n=\cd \: \vc D_n,  \quad  [\vc H]_t=0, \\
        [\vc{D}]_n=0,  \quad  [\vc E]_t=-\cd\: \vc{H}_t,
    \end{gather}
where rectangular brackets denote the difference of the field components at the boundary in the inner and outer domains. Inserting the expressions for the fields into these boundary conditions yields a system of equations
\begin{gather}
        R^3 \vc n \cdot \vc B_0 - \cd 2 \vc n\cdot \vc{d}_{\cd} = 2 \vc n \cdot \vc{m}_{\cd} \:,\\
        \frac{\vc n \times \vc B_0}{\mu}=-\frac{\vc n \times \vc{m}_{\cd}}{R^3} \:, \\
       \eps\left (\frac{2 \vc n \cdot \vc d_0}{\eps} + R^3 \: \vc E_0\cdot \vc n\right )=2 \vc n\cdot \vc{d}_{\cd} \:,\\        
        -\frac{\vc n \times \vc d_0}{\eps} + R^3 \vc n \times \vc E_0 + \cd R^3 \frac{\vc n \times \vc B_0}{\mu}=-\vc n\times \vc{d}_{\cd} \:. 
\end{gather}

Since the boundary conditions should be fulfilled at all points of the boundary, multiplications by $\vc n$ can be omitted and the system can be resolved as follows:
\begin{gather}
    \vc{d}_{\cd} = \frac{3(\mu +2)}{2\cd^2 \eps +(\eps+2)(\mu+2)} \vc d_0 \:, \label{eq:d_cd} \\
    \vc{m}_{\cd} = -\frac{6\cd}{2\cd^2 \eps +(\eps+2)(\mu+2)} \vc d_0 \:, \label{eq:m_cd}\\ 
    \vc E_0 = 2\frac{\vc d - \vc d_0}{\eps R^3} \:, \\
    \vc B_0 = -\mu \frac{\vc m}{R^3} \:.
\end{gather}

The induced dipole moments $\vc{d}_{\cd}$ and $\vc{m}_{\cd}$ are observable as they can be retrieved from the fields outside of the sphere. Equations~\eqref{eq:d_cd},\eqref{eq:m_cd} should be compared to the dipole moments found for an axion case in the same geometry~\cite{Seidov2023}:
\begin{gather}
    \vc{d}_{\c}=\frac{3(2+\mu)}{2\c^2\mu+(2+\mu)(2+\eps)} \vc d_0 \: , \label{eq:d_c}\\
    \vc{m}_{\c}=-\frac{3\chi\mu}{2\c^2\mu+(2+\mu)(2+\eps)}  \vc d_0 \: . \label{eq:m_c}
\end{gather}
A comparison of the results in equations \eqref{eq:d_cd}, \eqref{eq:m_cd} and \eqref{eq:d_c}, \eqref{eq:m_c} reveals that the induced electric dipole moments are practically the same up to the corrections quadratic in $\chi$ and $\cd$. In contrast, the induced magnetic moments are markedly different. 

As a next step, we assume that the effective material parameters in $\cd$ and $\chi$ case are connected to each other via the mapping Eqs.~\eqref{eq:eps_eff}-\eqref{eq:chi_eff}. Combining the mapping with Eq.~\eqref{eq:m_c}, we recover:
\begin{equation}
    \frac{m_\chi}{d_0}\longrightarrow\frac{3 \cd  \varepsilon }{2 (\mu +2)+\varepsilon  \left(\mu +2 \cd ^2+2\right)}\:.
\end{equation}
Importantly, this differs from Eq.~\eqref{eq:m_cd} by a factor of $-\eps/2$, so the results for the axion and the dual axion media are {\it nonequivalent} even if the respective material parameters are related to each other via the mapping.

To resolve this issue, we recall that the external sources should also be transformed accordingly, i.e. we need to compare the excitation of $\cd$ sphere by electric dipole $\vc{d}_0$ to the excitation of $\chi$ sphere by the combination of electric and magnetic dipoles. To demonstrate that, we solve the same problem but with a combination of electric and magnetic dipoles $\vc d_0$ and $\vc m_0$ inside an axion sphere (Fig.~\ref{fig:chi_tchi_d_dm}). We chose the fields inside and outside in the same form and the only difference is in the nonzero magnetic dipole field inside:
\begin{gather}
    \vc{E}_{\text{in}}=\frac{1}{\eps r^3}[3(\vc {n}\cdot\vc {d}_0)\vc {n}-\vc {d}_0]+\vc {E}_0\:, \\
    \vc{B}_{\text{in}}=\frac{1}{r^3} [3(\vc{n}\cdot\vc{m}_{0})\vc{n}-\vc{m}_{0}] + \vc B_0\:, \\
    \vc{E}_{\text{out}}=\frac{1}{r^3}[3(\vc{n}\cdot\vc{d})\vc{n}-\vc{d}] \:, \\
    \vc{B}_{\text{out}}=\frac{1}{r^3} [3(\vc{n}\cdot\vc{m})\vc{n}-\vc{m}] \:.
\end{gather}
The boundary conditions for the axion case are known to be
\begin{gather}
    [\vc{B}]_n=0, \quad [\vc{H}]_t=\c \vc{E}_t \:, \\
    [\vc{E}]_t=0, \quad [\vc{D}]_n=-\c \vc{B}_n \:,
\end{gather}
where the difference of the fields in inner and outer domains is denoted by the square brackets. Resolving the system of equations with respect to the unknown parameters yields
\begin{gather}
    \vc E_0 =\frac{\vc d_0- \eps \vc d}{\eps R^3}\:, \\
    \vc B_0 =2\frac{\vc m - \vc m_0}{R^3}\:, \\
    \vc m =3\frac{(\eps+2)\vc{m}_0-\c\mu\vc{d}_0}{2\c^2\mu+(\eps+2)(\mu+2)}\:, \\
    \vc d =\frac{6\c \vc{m}_0+3(2+\mu)\vc{d}_0}{2\c^2\mu+(\eps+2)(\mu+2)}\:.
\end{gather}

First, this result agrees with \eqref{eq:d_c} and \eqref{eq:m_c} if we set $\vc{m}_0=0$. Second, we analyze the situation when the sources are defined as
\begin{equation}
    \vc m_0=-\cd \vc d_1 \:, \quad \vc d_0=\frac{\mu}{\mu+\cd^2\eps} \vc d_1
\end{equation}
in agreement with the mapping, while the material parameters are expressed via Eqs.~\eqref{eq:eps_eff}-\eqref{eq:chi_eff}. We recover Eqs.~\eqref{eq:d_cd},\eqref{eq:m_cd} exactly, which illustrates the validity of the mapping.


Thus, we conclude that the dual axion medium with electric sources inside can be effectively discribed as an axion medium with both electric and magnetic sources. Since in the experimental setting the type of the introduced external sources is known in advance, it is possible to distinguish between $\c$ and $\cd$ responses. 

\section{IX. Solution for an oscillating dipole in $\chi$ or $\cd$ sphere}

In the section above we analyzed the scenario with static electric and magnetic dipoles exciting the structure. While being conceptually clear, this derivation misses one important aspect: the metamaterial we suggest features nonzero $\cd$ response only for the time-varying fields. Therefore, it is important to prove that $\c$ and $\cd$ responses can also be distinguished in the case of time-varying sources and fields. To that end, here we compute analytically the fields produced by the oscillating electric dipole exciting $\cd$ sphere. In our analysis we follow Ref.\cite{Seidov2023} employing multipole expansion technique.


Since the effects of spatially homogeneous $\cd$ are manifested only at the boundary, the fields inside and outside of a $\cd$ sphere can be described in the usual way. Specifically, the fields are expanded in terms of vector spherical harmonics~\cite{jackson} as follows:
\begin{gather}
        \vc{E}=\sum_{l,m}\left\lbrace\frac{i}{q}\nabla\times[a_E(l,m) N_l\vc{X}_{l,m}] - a_M(l,m) M_l \vc{X}_{l,m}\right\rbrace, \\
        \vc{B}=\sum_{l,m}\left\lbrace a_E(l,m) N_l\vc{X}_{l,m} + \frac{i}{q}\nabla\times[a_M(l,m) M_l \vc{X}_{l,m}]\right\rbrace.
\end{gather}
Here, $q=\omega/c$, $a_E(l,m)$ and $a_M(l,m)$ are multipole coefficients, $N_l$ and $M_l$ are linear combinations of spherical Bessel functions of the $l$-th order. Also,
\begin{equation}
    \vc{X}_{l,m}(\theta,\varphi)=\frac{1}{\sqrt{l(l+1)}}\hat{\vc{L}} Y_{l,m}(\theta,\varphi),
\end{equation}
is a VSH of degree $l$ and order $m$, $Y_{l,m}(\theta,\varphi)$ is a spherical harmonic of the same degree and order, and $\hat{\mathbf{L}}=-i{\bf r}\times\nabla$ is the angular momentum operator. Note also that in a spherical geometry surface normal vector $\mathbf n$ coincides with a normalized radius vector $\mathbf n \equiv \mathbf r / r$. We also use the identity
\begin{equation}
    \nabla\times[f(qr)\mathbf{X}_{l,m}]=\frac{1}{r}\underbrace{\frac{d}{dr}[r f(qr)]}_{D_r f(qr)} \underbrace{\mathbf{n}\times\mathbf{X}_{l,m}}_{\mathbf{Z}_{l,m}}+ i\sqrt{l(l+1)}\frac{f(qr)}{r}Y_{l,m}\mathbf{n}.
\end{equation}
Here we introduced for brevity a special derivative operator $D_r$ and an auxiliary VSH $\mathbf{Z}_{l,m}$. Note, that the first term is tangential to the surface of the sphere and orthogonal to $\mathbf X_{l,m}$, and thus $\mathbf X_{l,m}, \mathbf Z_{l,m}$ form a full basis in the tangential space for each $(l,m)$. The advantage of such multipole decomposition is that it allows to reduce the solution of Maxwell's equations to an algebraic problem of finding the expansion coefficients $a_E(l,m)$ and $a_M(l,m)$, and since $\mathbf X_{l,m}, \mathbf Z_{l,m}$ and ${\bf n}\,Y_{l,m}$ are linearly independent for each $(l,m)$ and between each other, if the boundary conditions are also linear, we can find expansion coefficients separately for each set of $l,m$ indices. This expansion is called multipole, because every VSH of the $l$-th order corresponds to the respective term in the multipole expansion~\cite{jackson}.

As we consider the excitation of the sphere by the dipole source, we include the terms of the multipole expansion with $l=1$, while higher-order multipole harmonics vanish. The indices $I$ and $O$ correspond to the fields inside and outside of the sphere, respectively. Then we write   \begin{gather}
        \sqrt{\varepsilon}\mathbf{E}_I=\frac{i}{Q}\nabla\times[(A h_1^{(1)}(Qr) + B j_1(Qr))\mathbf{X}_{10}(\theta,\varphi)]- 
        \tilde B j_1(Qr) \mathbf{X}_{10}(\theta,\varphi), \label{E_I_expansion} \\
        \sqrt{\mu}\mathbf{H}_I=(A h_1^{(1)}(Qr) + B j_1(Qr))\mathbf{X}_{10}(\theta,\varphi) +   \frac{i}{Q}\nabla\times[\tilde B j_1(Qr) \mathbf{X}_{10}(\theta,\varphi)], \label{H_I_expansion} \\
        \mathbf{E}_O=\frac{i}{q}\nabla\times[a^{\tilde \chi}_E h_1^{(1)}(qr)\mathbf{X}_{10}(\theta,\varphi)] -  a_M^{\tilde \chi} h_1^{(1)}(qr) \mathbf{X}_{10}(\theta,\varphi), \label{E_O_expansion} \\
        \mathbf{H}_O=a^{\tilde \chi}_E h_1^{(1)}(qr)\mathbf{X}_{10}(\theta,\varphi) + \frac{i}{q}\nabla\times[ a_M^{\tilde \chi} h_1^{(1)}(qr) \mathbf{X}_{10}(\theta,\varphi)].  \label{H_O_expansion}
\end{gather}
The radial functions for the fields inside and outside of the sphere are spherical Bessel function $j_1(qr)$ and spherical Hankel function $h_1^{(1)}(qr)$, respectively. Here, $q=\omega/c$, $Q=\sqrt{\varepsilon\mu}q$. The electric dipole expansion coefficient $A=-id_0\varepsilon\mu^{3/2}q^3\sqrt{8\pi/3}$ is expressed via the dipole moment of the external source $d_0$, while $B, \tilde B, a^{\tilde \chi}_E$ and $a_M^{\tilde \chi}$ are the unknown coefficients. The choice of the radial functions is dictated by the physics of the problem: Hankel functions for $e^{-i\omega t}$ time convention correspond to the outgoing spherical waves, while Bessel function is regular at the origin and corresponds to the boundary-induced field inside the sphere.

For monochromatic fields it is sufficient to satisfy the boundary conditions for the tangential components, while the conditions for the normal components are fulfilled automatically. The tangential boundary conditions read
\begin{gather}
    \mathbf{H}_{1t}=\mathbf{H}_{2t}, \label{bc:cd_H}\\
    \mathbf{E}_{1t}-\mathbf{E}_{2t}=\cd \mathbf{H}_{1t}. \label{bc:cd_E}
\end{gather}
The tangential space of the sphere at each point is spanned by the vectors $\mathbf{X}_{10}$ and $\mathbf{Z}_{10}$. Therefore, each of the boundary conditions \eqref{bc:cd_H}, \eqref{bc:cd_E} yields two scalar equations. In total, at the boundary $r=R$ we obtain four linear equations with four unknowns $B, \tilde B, a^{\tilde \chi}_E$ and $a_M^{\tilde \chi}$ 
\begin{gather}
    a^{\tilde \chi}_E h_1^{(1)}(qR)=\frac{A}{\sqrt{\mu}} h_1^{(1)}(QR) + \frac{B}{\sqrt{\mu}} j_1(QR), \\
    a_M^{\tilde \chi} D_R h_1^{(1)}(qR)= \frac{\tilde B}{\mu\sqrt{\varepsilon}} D_R j_1(QR), \\
    -\frac{\tilde B}{\sqrt{\varepsilon}} j_1(QR) + a_M^{\tilde \chi} h_1^{(1)}(qR)=\cd a^{\tilde \chi}_E h_1^{(1)}(qR), \\
    \frac{A}{\varepsilon \sqrt{\mu}}D_R h_1^{(1)}(QR) + \frac{B}{\varepsilon \sqrt{\mu}} D_R j_1(QR) - a^{\tilde \chi}_E D_R h_1^{(1)}(qR)=\cd  a_M^{\tilde \chi} D_R h_1^{(1)}(qR).
\end{gather}
We introduce the notations
\begin{gather}
    \alpha=\frac{j_1(Q R)}{h_1^{(1)}(q R)}, \\
    \beta=\frac{D_R j_1(Q R)}{D_R h_1^{(1)}(q R)}, \\
    \delta=\frac{h_1^{(1)}(Q R)}{h_1^{(1)}(q R)}, \\
    \gamma=\frac{D_R h_1^{(1)}(Q R)}{D_R h_1^{(1)}(q R)}.
\end{gather}
This allows us to rewrite our system in the matrix form
\begin{gather}
    \begin{pmatrix}
     -\frac{\alpha }{\sqrt{\mu}} & 0 & 1 & 0 \\
    0 & -\frac{\beta }{\sqrt{\varepsilon} \mu} & 0 & 1 \\
    0 & -\frac{\alpha }{\sqrt{\varepsilon}} & \cd & 1 \\
    \frac{\beta }{\varepsilon \sqrt{\mu}} & 0 & -1 & \cd \\
    \end{pmatrix} 
    \begin{pmatrix}
        B \\ \tilde B \\ a^{\tilde \chi}_E \\ a_M^{\tilde \chi}
    \end{pmatrix}=
    \begin{pmatrix}
        \frac{A \gamma }{\sqrt{\mu}} \\ 0 \\ 0 \\ -\frac{A \delta }{\varepsilon \sqrt{\mu}}
    \end{pmatrix}.
\end{gather}

The solution of such system reads 
\begin{gather}
    B = \frac{A \left(\gamma  \varepsilon \left(\beta -\alpha  \mu+\beta  \cd^2\right)+\delta  (\alpha  \mu-\beta )\right)}{\alpha  \varepsilon \left(\alpha  \mu-\beta  \left(\cd^2+1\right)\right)+\beta  (\beta -\alpha  \mu)}, \\ 
    \tilde B = \frac{A \sqrt{\varepsilon \mu} \cd (\alpha  \delta -\beta  \gamma )}{\alpha  \varepsilon \left(\alpha  \mu-\beta  \left(\cd^2+1\right)\right)+\beta  (\beta -\alpha  \mu)}, \\
    a^{\tilde \chi}_E =  \frac{A (\alpha  \mu-\beta ) (\alpha  \delta -\beta  \gamma )}{\sqrt{\mu} \left(\alpha  \varepsilon \left(\alpha  \mu-\beta  \left(\cd^2+1\right)\right)+\beta  (\beta -\alpha  \mu)\right)}, \label{eq:atChiE}\\
    a_M^{\tilde \chi} = \frac{A \beta  \cd (\beta  \gamma -\alpha  \delta )}{\sqrt{\mu} \left(\alpha  \varepsilon \left(\beta -\alpha  \mu+\beta  \cd^2\right)+\beta  (\alpha  \mu-\beta )\right)}. \label{eq:atChiM}
\end{gather}
Since the expansion coefficients of the fields outside the sphere are related to the effective dipole moments in the same manner as the coefficients inside the sphere are related to the dipole moments of the external source, we evaluate the ratios of the dipole moments as 
\begin{gather}
    \frac{a^{\tilde \chi}_E}{A} \varepsilon\mu^{3/2} =\frac{d}{d_0},\\
    \frac{a_M^{\tilde \chi}}{A} \varepsilon\mu^{3/2}=\frac{m}{d_0}.
\end{gather}
For consistency, we check that the static limit of our solution coincides with an analytical result obtained in the previous section. The static limit of our expression is obtained via Taylor series expansion and yields
\begin{gather}
    \frac{m}{d_0}\bigg|_{q=0}=-\frac{6 \cd}{\left(\varepsilon+2\right) \left(\mu+2\right)+2\eps\cd^2}, \\
    \frac{d}{d_0}\bigg|_{q=0}=\frac{3(\mu+2)}{\left(\varepsilon+2\right) \left(\mu+2\right)+2\eps\cd^2}.
\end{gather}
This coincides with Eqs.~\eqref{eq:d_cd},\eqref{eq:m_cd} obtained previously.

\begin{figure}
    \centering
    \includegraphics[scale=0.5]{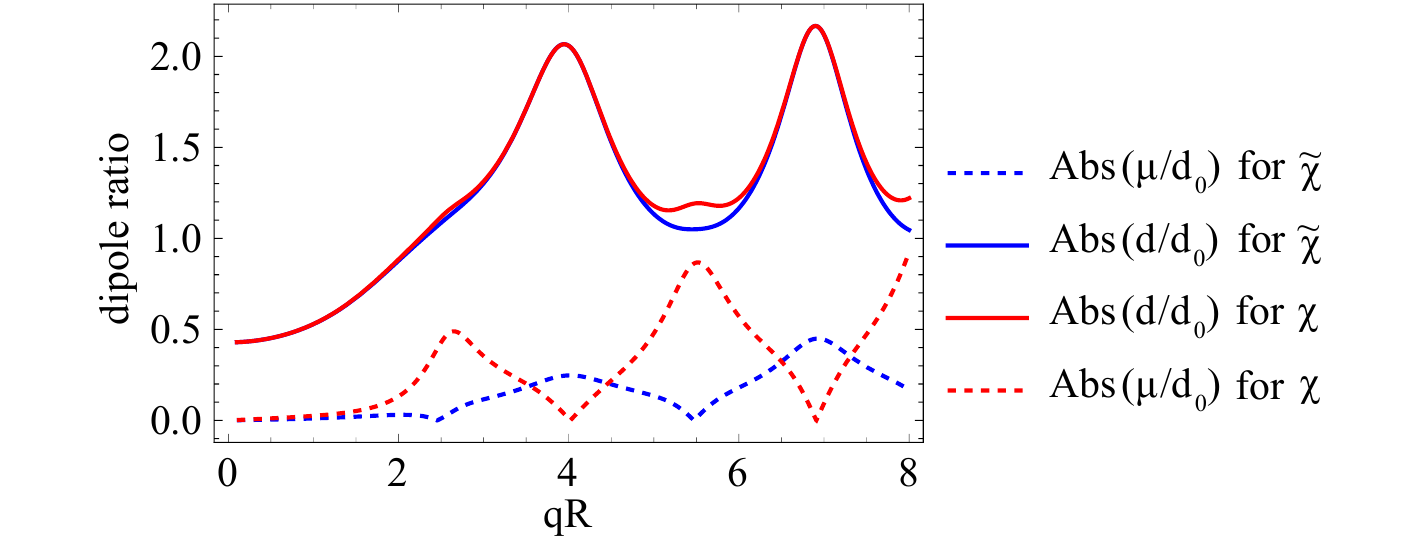}
    \caption{The ratios of the effective dipole moment to the dipole moment of the introduced source computed for $\c$ and $\cd$ spheres. For dual axion case the following parameters are chosen: $R=1$, $\varepsilon=5$, $\mu=1$, $\cd=0,03 qR$. For axion case, the respective parameters $\varepsilon_{\text{eff}}, \mu_{\text{eff}}, \c_{\text{eff}}$ are calculated via $\varepsilon,\mu,\cd$ using the mapping Eqs.~\eqref{eq:eps_eff},\eqref{eq:mu_eff},\eqref{eq:chi_eff}.}
    \label{fig:enter-label}
\end{figure}

The same calculation has been performed in Ref.~\cite{Seidov2023} for the axion sphere. The coefficients of multipole expansion in this case are
\begin{gather}
    a_E^\c=\frac{A (\alpha  \mu-\beta ) (\alpha  \delta -\beta  \gamma )}{\sqrt{\mu} \left(\alpha  \varepsilon (\alpha  \mu-\beta )+\beta  \left(\beta -\alpha  \mu \left(\c^2+1\right)\right)\right)}, \\
    a_M^\c=\frac{\alpha  A \sqrt{\mu} \c (\alpha  \delta -\beta  \gamma )}{\alpha  \varepsilon (\beta -\alpha  \mu)+\beta  \left(\alpha  \mu \left(\c^2+1\right)-\beta \right)}.
\end{gather}
The other two coefficients are irrelevant to our purposes. Comparing this result to Eq.~\eqref{eq:atChiE} and \eqref{eq:atChiM}, we observe that in the first order in $\c$ and $\cd$ coefficients $a_E^\chi$ and $a_E^{\tilde \chi}$ coincide and their difference is of the second order at most. Since the typical values of $\chi$ are well below unity, the dipole ratios $d/d_0$ practically coincide. On the other hand, $a_M^\chi$ and $a_M^{\tilde \chi}$ differ already in the first order in $\c$ and $\cd$ as illustrated in Fig.~\ref{fig:enter-label}.

Inspecting the obtained results, we observe that the induced effective magnetic dipole moment features the  nontrivial dependence on frequency. At certain frequencies the induced magnetic dipole vanishes which corresponds to the non-radiative current configurations. For the case of  axion sphere, the positions of these zeros are dictated by the zeros of the Bessel function $j_1(\sqrt{\varepsilon\mu} q R)$, which can immediately derived from the fact that  $a_M^\c$ is proportional to $\alpha$. At the same time, $a_M^{\tilde \chi}$ is proportional to $\beta$. Therefore, in the dual axion sphere the non-radiative current configurations appear at zeros of $D_R j_1(\sqrt{\varepsilon \mu} q R)$. 

Thus, frequency dependence of the induced magnetic dipole moment allows to clearly distinguish $\c$ and $\cd$ responses in the dynamic case as well.


\bibliography{AxionLib}